\documentclass[aps,pra,tightenlines,10pt,notitlepage,nofootinbib,twocolumn,superscriptaddress]{revtex4-2}
\usepackage{amsmath,epsfig,amssymb}
\usepackage{bbm}
\usepackage{bm}

\usepackage{afterpage}

\usepackage{times}
\usepackage{amsthm}
\usepackage{amsmath}
\usepackage{amsfonts}
\usepackage{comment}
\usepackage{color}
\usepackage[caption=false]{subfig}
\usepackage{multirow}
\usepackage[normalem]{ulem}
\usepackage{qcircuit}
\usepackage{braket}
\usepackage{amsfonts,amssymb,dsfont}
\usepackage{graphicx}

\usepackage{inputenc}
\usepackage{dsfont}
\usepackage{array}
\usepackage{tabularx}
\usepackage[ruled,vlined]{algorithm2e}

\usepackage[acronym,shortcuts]{glossaries}
\usepackage{hyperref}
\hypersetup{
    colorlinks=true,       
    linkcolor=red,          
  citecolor=magenta,        
    filecolor=magenta,      
    urlcolor=cyan,          
    runcolor=cyan
}

\newcolumntype{C}[1]{>{\centering\arraybackslash}p{#1}}

\begin{document}
\title{Entanglement Scaling and Problem Structure in Quantum Approximate and Adiabatic Optimization Algorithms}
\author{Georgios Arapantonis}
\affiliation{William H. Miller III Department of
Physics and Astronomy, Johns Hopkins University, Baltimore, Maryland 21218, USA}

\author{Paraj Titum}
\affiliation{William H. Miller III Department of
Physics and Astronomy, Johns Hopkins University, Baltimore, Maryland 21218, USA}
\affiliation{Johns Hopkins University Applied Physics Laboratory, Laurel, Maryland 20723, USA}

\author{Gregory Quiroz}
\affiliation{William H. Miller III Department of
Physics and Astronomy, Johns Hopkins University, Baltimore, Maryland 21218, USA}
\affiliation{Johns Hopkins University Applied Physics Laboratory, Laurel, Maryland 20723, USA}

\begin{abstract}
Entanglement is widely regarded as a key resource underlying the power of quantum algorithms and their potential to achieve quantum advantage. With the emergence of variational quantum algorithms, however, questions have arisen regarding how entanglement relates to problem structure and algorithmic performance in near-term quantum applications. Here, we examine this relationship through the Quantum Approximate Optimization Algorithm (QAOA), a specific class of variational algorithms, applied to the MaxCut problem. We show that suboptimal variational parameter training can significantly modify the observed entanglement profile, obscuring its scaling behavior. By employing a high-performance optimizer, we find empirical evidence that QAOA exhibits entanglement scaling consistent with that of fermionic Gaussian states (up to a scaling factor) across a broad range of MaxCut instances. We further compare these results with adiabatic quantum computation, observing annealing-schedule-dependent entanglement profiles whose scaling behavior differs markedly from that of QAOA. Together, these findings provide new insight into how entanglement manifests in and distinguishes these two algorithmic paradigms, highlighting its connection to both computational performance and problem structure.
\end{abstract}
\maketitle

\section{Introduction}

Variational quantum algorithms (VQAs) have emerged as a leading approach to quantum computing on near-term quantum devices~\cite{Cerezo_VQA_Review_2021, quantum_optimization_review_2023,Mohseni_quantum_computer_scaling_2025}. In a typical application, a VQA implements a variational ansatz to approximate the ground state of a quantum Hamiltonian and its corresponding energy. VQAs offer two main advantages \cite{Cerezo_VQA_Review_2021} that make them a favorable framework for near-term quantum processors. First, their learning-based approach provides a flexible strategy that collectively addresses the constraints of different hardware architectures~\cite{Cerezo_VQA_Review_2021}. Second, VQAs can exhibit inherent robustness to noise and generally require lower circuit depth circuits compared to fault-tolerant approaches for finding ground states \cite{Cerezo_VQA_Review_2021,Langfitt_Paremeter_Transferability_2023,Hunag_Noise_Robust_VQA_2022}.
The versatility of VQAs allows them to tackle a wide range of tasks, and it has been shown that they can even be employed for universal quantum computing \cite{Biamonte_universal_quantum_computation_2021}. Indeed, VQAs have found applications in quantum simulations \cite{Yuan2_VQAs_simulation_2019,Li_VQA_2017,McLachlan_Schrodinger_VQA_1964,McArdle_VQA_ImaginaryTime_2019,Endo_VQA_2020,Yao_ADAPT_VQA_2021,Zhang_ADAPT_Simulation_2023}, the approximation of molecular ground states \cite{Peruzzo_VQE_Photons_2014,Nakanishi_VQE_2019,Parrish_VQE_2020,McClean_QO_2021,Wang_VQE_2019,Wang_EstimationRuntime_2021}, and combinatorial optimization problems \cite{Moll_opt_2018,Wang_FermionicQAOA_2018,Romero_UCC_2019,Wurtz_MaxCut_QAOA_2021,Wecker_Opt_2016,Khairy_VQA_2020}.

The quantum approximate optimization algorithm (QAOA) is a well-studied instance of VQAs designed to approximately solve classical combinatorial optimization problems~\cite{Farhi_QAOA_2014}. 
$p$-QAOA involves $p$ parameterized layers, each designated by variational parameters. In its canonical form, every layer contains one mixer term and cost evolution that work together to boost the amplitude of solution states. The $2p$ variational parameters---one for each mixer/phase term per layer---are optimized using classical algorithms to minimize a cost function encoding the computational problem. Numerous studies have demonstrated QAOA's potential on near-term devices for determining the ground state of Ising Hamiltonians \cite{Shaydulin_QAOA_Np200_2023,Google_QAOA_NonPlanarGraphs_2021,Leontica_SK_2024,Pogano_QAOA_TrappedIon_2020,Lykov_QAOA_quantum_advantage_2023,Shaydulin_QAOA_advantage_2024,He_QAOA_advantage_2025}. Notably, there exists an inherent relationship between QAOA and adiabatic quantum computation (AQC)~\cite{Farhi_AQC_2000, Lidar_AQC_2018}, as the alternating structure of QAOA layers can be viewed as a time-discretized, variational version of continuous-time AQC. That connection has inspired translations between the two algorithms that assist with investigations of QAOA dynamics and variational parameter training~\cite{Zhou2020PRX,Sack_QAInitialization_2021,Wurtz_Love_2022_counterdiabaticity,Headley_QAOA_AQC_2022,Chen2022CounderdiabaticQAOA}. 

Recently, numerous studies have investigated entanglement in QAOA \cite{DupontEntanglementPerspective2022,DupontCalibratingClassicalHardness2022,Sreedhar_QAOAEntMPS_2022, Miki_ScalingFunctionQAOA_2025} and, more broadly, in VQAs \cite{Woitzik_VQE_2020,Wiersema_EntOpt_2020,Usman2024CalibratingRoleOfEntanglement,Chen_2022_ADAPTQAOA,Sim_ExpressCircuits_2019,Wiersema_MeausrementEnt_2023,Wiersema_EntOpt_2020} and AQC \cite{Latorre_AQCEntPhaseTransition_2004,Orus_EntComplexity_May2004,Bauer_EntResourceAQC_2015,Batle_CorrelationsAQC_2016,Schutzhold_AQCPhaseTransitions_2006}. Entanglement is widely recognized as a key resource in quantum systems and quantum algorithms that distinguishes them from their classical counterparts \cite{NielsenChuang2010,Horodecki_quantum_entanglement_2009}. For example, the growth of entanglement, together with the entanglement spectrum, has been employed to characterize the properties of quantum states in strongly correlated many-body systems \cite{Laflorencie_entanglement_condensed_matter_2016}. Previous studies examining entanglement in AQC have further linked features of entanglement growth to the system undergoing a quantum phase transition \cite{Latorre_AQCEntPhaseTransition_2004,Schutzhold_AQCPhaseTransitions_2006}. Recent analytical and numerical works have also developed tools to investigate the entanglement dynamics of systems out of equilibrium, particularly following a quantum quench~\cite{Xu_LocalityScrambling_2019,Nahum_OperSpreading_2018}. VQAs have been examined through the lens of quench dynamics, as the alternating application of mixer and cost Hamiltonians in algorithms such as QAOA can be viewed as a sequence of controlled quenches~\cite{Nagano_Schwinger_VQAs_2023}. A recent study exploited a quantum quench as a shortcut to adiabaticity for preparing ground states using quantum algorithms \cite{Lukin_quantum_quench_dynamics_2024}. Thus, methods developed to study entanglement dynamics in quench protocols offer a natural and promising framework for investigating entanglement in VQAs.

In recent years, systematic investigations of the growth of entanglement in QAOA have emerged. For random QAOA circuits \cite{DupontEntanglementPerspective2022} (i.e., those with randomly chosen variational parameters), it has been found that the entanglement at the end of the circuit increases monotonically with circuit depth and converges to the Page value, the average entanglement entropy of a subsystem of a quantum system that is in a random pure state \cite{PageValueRandom1993}. In contrast, for optimized QAOA circuits, it is found that there is an entanglement barrier between the initial superposition state and the final solution state, which scales as a volume law with the system size or number of qubits \cite{DupontEntanglementPerspective2022}. In addition, it has been shown that different mixer Hamiltonians have varying entangling and disentangling abilities. For example, it was shown that ADAPT-QAOA---a protocol for adaptively training and ``growing" a QAOA ansatz \cite{Zhu_ADAPT_QAOA_2022}---offers greater flexibility in both generating and removing entanglement between qubits, whereas the traditional QAOA ansatz cannot remove excess entanglement efficiently~\cite{Chen_2022_ADAPTQAOA}. 

Moreover, the role of entanglement in guiding the algorithm toward the solution remains subtle. Recent studies \cite{Chen_2022_ADAPTQAOA,Sreedhar_QAOAEntMPS_2022} have found that greater entanglement does not necessarily lead to better performance. Nevertheless, alternative metrics such as tripartite mutual information show a positive correlation with the complexity of quadratic unconstrained binary optimization instances~\cite{Qian_Scrambling_2024}. In addition, previous work~\cite{DupontCalibratingClassicalHardness2022,Miki_ScalingFunctionQAOA_2025} has associated the difficulty of simulating QAOA with matrix product states (MPSs) with the entanglement generated during the evolution, since the efficiency of MPS representations depends on the amount and distribution of bipartite entanglement in the state. However, Ref.~\cite{Usman2024CalibratingRoleOfEntanglement} reports that this behavior is sensitive to both circuit depth and graph structure. Despite the progress in characterizing entanglement growth in QAOA, the relationship between the entanglement dynamics, algorithmic performance, and the specifications of the computational problem remains an open question.

In this work, we elucidate important details about the relationship between these characteristics by investigating the entanglement profile of QAOA as a function of problem structure. This is accomplished using the MaxCut problem---a widely studied combinatorial optimization problem in graph theory with numerous applications \cite{Commander2009MaxCutReview}. We first examine the impact of the classical training strategy on the qualitative features of the QAOA entanglement profile, providing insights into the interplay between algorithmic performance and entanglement growth. We show that different training protocols yield differences in the entanglement profiles for complete graphs, highlighting the significance of the classical training component in interpreting entanglement dynamics. 

We further characterize entanglement spreading in QAOA by studying entanglement as a function of graph partitions. Recent work \cite{DupontEntanglementPerspective2022,Usman2024CalibratingRoleOfEntanglement} has focused primarily on the scaling of entanglement for balanced bipartitions, where each subsystem contains half of the qubits. While balanced cuts are often sufficient for characterizing the asymptotic scaling of entanglement, they provide only a partial view of the entanglement structure. Examining unbalanced bipartitions of $N_1$ and $N-N_1$ qubits probes entanglement across a range of subsystem sizes, revealing how correlations are distributed throughout the system and providing a more complete characterization of entanglement growth. This perspective has proven particularly useful in studies of nonequilibrium many-body dynamics following a quantum quench~\cite{Nahum_Ent_Growth_Quench_2017}, a setting closely related to the circuit dynamics underlying QAOA, as it reveals universal structure behind entanglement growth. Through numerical simulations, we analyze the entanglement of unbalanced bipartitions required to solve the MaxCut problem, normalized by the required entanglement of the balanced bipartition. We find that, when normalized by the entanglement of the balanced bipartition, the entanglement of unbalanced bipartitions collapses onto a single scaling curve across multiple system sizes. We then show that this universal scaling behavior can be understood through a correspondence with fermionic Gaussian states (FGSs)~\cite{Surace_FGS_notes_2022}.

The structure of the computational problem is also shown to be reflected in the entanglement generated by both QAOA and AQC, the continuous-time counterpart of QAOA. Most previous studies have focused on fixed graph instances and entanglement across balanced bipartitions, limiting the ability to identify systematic relationships between entanglement and graph properties. Here, we instead examine graph ensembles with varying edge density and relate this structural parameter to the entanglement required to solve MaxCut. For QAOA, we provide numerical evidence that the maximum generated entanglement obeys a simple scaling relation with edge density in the large-density regime, independent of the specific graph instance. For AQC, we show that the scaling relation of the entanglement barrier as a function of graph features depends on the chosen annealing path. Together, these results establish a direct connection between problem structure and the entanglement profiles of both QAOA and AQC.

The remainder of the manuscript is organized as follows. In Sec.~\ref{Section:Background}, we present the necessary background on QAOA and AQC, as well as the entanglement metric used in this work. In Sec.~\ref{Section:Methods}, we describe the classical training scheme used to train the QAOA, the methodology for calculating entanglement entropy, and the specifications of the problem instance datasets. Sec.~\ref{Section:optimization_strategy_effect_qaoa} and Sec.~\ref{Section:problem_structure_ent_scaling_QAOA} present the main results for QAOA, while Sec.~\ref{Section:impact_annealing_schedule_AQC} and Sec.~\ref{Section:problem_structure_ent_scaling_AQC} provide the analysis for AQC. Finally, in Sec.~\ref{Section:Summary}, we summarize our findings, discuss their implications, and outline possible directions for future research.

\section{Background}
\label{Section:Background}

\subsection{MaxCut Problem}
\label{Section:MaxCut}

The MaxCut problem is defined on a graph $ G=(V, E) $ with a set of vertices $ V $ and a set of edges $ E $. Each edge $ (i, j) \in E $ has an associated weight $ w_{ij} $. The MaxCut problem involves partitioning the vertices into two disjoint sets such that the total weight of the edges between the sets is maximized. Formulations of the problem can be described via a classical objective function $ C(x)$. Potential solutions to the problem are encoded as $N $-bit binary strings $ x = x_1x_2 \ldots x_N $, with $ x_i \in \{0,1\} $, $i=1,\dots,N $. The classical objective function $ C(x) $ can be mapped to a corresponding quantum cost Hamiltonian $H_C $ by associating the binary variables $x_j$ with the $\pm1$ eigenstates of the Pauli-Z spin operator $\sigma^z_j$ acting on qubit $j$, via $\sigma^z_j \ket{x_j}=(-1)^{x_j}\ket{x_j}$. The structure of $H_C$ is such that the ground state of the Hamiltonian is the solution to the computational problem.

Executing the translation procedure above results in the cost Hamiltonian 
\begin{equation} 
H_C = \sum_{( i,j )\in E} w_{ij} \sigma^z_i \sigma^z_j.
\label{Eq:HamiltonianMaxCutDegenerate}
\end{equation}
 By construction, the cost Hamiltonian is diagonal in the computational basis. The ground state manifold possesses a $\mathbb{Z}_2$ symmetry, and thus, includes a minimum of two states $\ket{x}$ and $ \ket{\bar{x}} $, where $\ket{\bar{x}} = \left( \bigotimes^N_{j=1} \sigma^x_j\right) \ket{x}$. Provided the distribution of weights $w_{ij}$ is chosen such that no additional degeneracies are present, the ground state of Eq.~\eqref{Eq:HamiltonianMaxCutDegenerate} is 
\begin{equation}
    \ket{\Psi} = \frac{1}{\sqrt{2}} (\ket{x} + e^{i \phi} \ket{\bar{x}}),
\label{Eq:SolutionDegenerateMaxCut}
\end{equation}
with $\phi$ being an arbitrary phase. 

\subsection{QAOA}
QAOA is a hybrid quantum-classical algorithm designed to find approximate solutions to combinatorial optimization problems, like the MaxCut problem~\cite{Farhi_QAOA_2014}. The standard $p$-QAOA takes the form (assuming $\hbar=1$)
\begin{equation}
\ket{\psi_p(\vec{\boldsymbol{\beta}}_p, \vec{\boldsymbol{\gamma}}_p)} = \left ( \prod_{j=1}^{p} e^{-i \beta_j H_M} e^{-i \gamma_j H_C} \right ) \ket{\Psi_0},
\label{Eq:VariationalStatevector}
\end{equation}
where the system is initialized in the state $\ket{\Psi_0}=\ket{+}^{\otimes N}$, with $\ket{+}$ being the $+1$ eigenstate of the $\sigma^x$ Pauli operator. The subsequent evolution is generated by two Hamiltonians: the mixer Hamiltonian $H_M$ and the cost Hamiltonian. Here, we use the canonical choice $H_M = \sum_{j=1}^{N} \sigma^x_j$. The computational problem---taken as the MaxCut problem in this work---specifies the cost Hamiltonian. Time evolution of the mixer and cost are parameterized by the variational parameters $ \vec{\boldsymbol{\beta}}_p = (\beta_1, \beta_2, \ldots, \beta_p) $ and $ \vec{\boldsymbol{\gamma}}_p = (\gamma_1, \gamma_2, \ldots, \gamma_p)$, respectively.

For given parameters $ \vec{\boldsymbol{\beta}}_p $ and $ \vec{\boldsymbol{\gamma}}_p $, the expectation value of $ H_C $ is then calculated via
\begin{equation}
F_p(\vec{\boldsymbol{\beta}}_p, \vec{\boldsymbol{\gamma}}_p) = \langle \psi_p(\vec{\boldsymbol{\beta}}_p, \vec{\boldsymbol{\gamma}}_p) | H_C | \psi_p(\vec{\boldsymbol{\beta}}_p, \vec{\boldsymbol{\gamma}}_p) \rangle.
\label{Eq:VariationalCostFunction}
\end{equation}
Classical training methods are employed to determine optimized variational parameters $( \vec{\boldsymbol{\beta}}^*_p , \vec{\boldsymbol{\gamma}}^*_p)$ that minimize the expectation value of the Hamiltonian $H_C$. The algorithmic performance is evaluated using the approximation ratio
\begin{equation}
    r = \frac{C_{max} - F_p(\vec{\boldsymbol{\beta}_p}, \vec{\boldsymbol{\gamma}}_p)}{C_{max} - C_{min}},
\label{Eq:ApproximationRatio}
\end{equation}
which, during an optimization run, reaches its maximum value for the optimized parameters. Note that $ C_{\min} $ and $ C_{\max} $ represent the global minimum and maximum eigenvalues of the cost Hamiltonian $ H_C $. The approximated solution is optimal when $ r = 1 $, which is theoretically achievable as $ p \rightarrow \infty $.

\subsection{Adiabatic Quantum Computing}
AQC implements quantum algorithms by guiding a many-body quantum system through a controlled adiabatic evolution. Specifically, it solves the computational problem by slowly driving the system to the ground state of a quantum Hamiltonian, which encodes the solution to the computational problem. The time-dependent Hamiltonian $H(s)$ is typically expressed as:
\begin{equation}
H(s) = -\left(1 - f(s)\right) H_M + f(s) H_C, 
\label{Eq:TimeDependentHamiltonian}
\end{equation}
where $s=t/T$ is the normalized time and $0\leq t \leq T$. $f(s)$ is a monotonic interpolation function satisfying $f(0) = 0$ and $f(1) = 1$. The system begins in the easily prepared ground state $\ket{\Psi(0)} = \ket{\Psi_0}$ of $-H_M$, and evolves to the ground state of $H_C$ according to the Schr\"{o}dinger equation. The final state is given by
\begin{equation}
\label{Eq:Final_State_AQC}
    \ket{\Psi(T)} = \mathcal{T} \exp\left(-iT\int_0^1 H(s)\, ds\right) \ket{\Psi_0},
\end{equation}
where $\mathcal{T}$ denotes the time-ordering operator.

The success of AQC relies on the adiabatic theorem, which states that if the Hamiltonian changes sufficiently slowly, the system will remain in its instantaneous ground state throughout the evolution \cite{Farhi_AQC_NP_2001}. Quantifying the slow rate of change is a subtle problem addressed by various versions of the adiabatic theorem. Broadly speaking, the adiabatic theorem bounds the runtime $T$ using the minimum spectral gap $\Delta_{\min}$ between the ground and first excited states during the evolution~\cite{Jansen_BoundsAQC_2007}. For a comprehensive review, see Ref.~\cite{Lidar_AQC_2018}.

There is an immediate connection between AQC and QAOA. Specifically, QAOA can be interpreted as a digital approximation of the continuous adiabatic evolution in AQC, where the total runtime is discretized into $p$ steps, such that $T=p\Delta t$. This correspondence becomes explicit when discretizing the time evolution operator in Eq.~\eqref{Eq:Final_State_AQC} using the first-order Suzuki-Trotter expansion~\cite{Susuzki_Approx_1990}:
\begin{equation}
    \mathcal{T}e^{-iT\int^1_0 H(s) \, ds} \approx \prod_{j=1}^{p} e^{-i \beta_j H_M} e^{-i \gamma_j H_C},
\label{Eq:AdiabaticTheoremOriginal}
\end{equation}
where the parameters $\beta_j = -(1 - f(s_j))\, \Delta t$, $\gamma_j=f(s_j) \Delta t$ define the Trotterized evolution at discrete times $s_j = j/p$. While the digitized annealing schedule can be used to define the QAOA evolution, it is common to consider more general parameterizations. Nevertheless, it is worth noting that the AQC schedule can be used to initialize the variational parameters \cite{Sack_QAInitialization_2021}, or assist in reducing the dimension of the parameter space \cite{Sakai_LinearlityQAOAParams_2025,Wu2024_AdiabaticPassageParameters_2024}.

\subsection{Entanglement Entropy}

Consider a quantum system composed of $N$ qubits, whose state resides in a Hilbert space $\mathcal{H}$, with $\dim(\mathcal{H}) = 2^N$. The QAOA states defined in Eq.~\eqref{Eq:VariationalStatevector} are pure, meaning the global density matrix $\rho$ satisfies $\rho^2 = \rho$, and thus the total system has von Neumann entropy $S = -\mathrm{Tr}(\rho \ln \rho)=0$. However, this does not imply the absence of quantum correlations among subsystems. When a subset of qubits is traced out, the resulting reduced state is generally mixed, reflecting the presence of entanglement between the remaining and discarded subsystems.

To quantify such bipartite entanglement, we consider a decomposition of the system into two subsystems with Hilbert spaces $\mathcal{H}_1$ and $\mathcal{H}_2$, such that $\mathcal{H} = \mathcal{H}_1 \otimes \mathcal{H}_2$. For a pure state $|\psi\rangle \in \mathcal{H}$, the Schmidt decomposition guarantees the existence of orthonormal bases $ \{ |i_1\rangle \} $ for $\mathcal{H}_1$ and $\{ |i_2\rangle \}$ for $\mathcal{H}_2$, such that
\begin{equation*}
    |\psi\rangle = \sum_i \lambda_i |i_1\rangle \otimes |i_2\rangle.
\end{equation*}

The non-negative real coefficients $\{ \lambda_i \}$, known as Schmidt coefficients, satisfy $\sum_i \lambda_i^2 = 1$. The reduced density matrices of the two subsystems are then given by
\begin{align*}
    \rho^{\mathcal{H}_1} &= \mathrm{Tr}_{\mathcal{H}_2}(|\psi\rangle\langle\psi|) = \sum_i \lambda_i^2 |i_1\rangle\langle i_1|, \\
    \rho^{\mathcal{H}_2} &= \mathrm{Tr}_{\mathcal{H}_1}(|\psi\rangle\langle\psi|) = \sum_i \lambda_i^2 |i_2\rangle\langle i_2|,
\end{align*}
where the reduced density matrices share the same set of eigenvalues $\{ \lambda_i^2 \}$. As a result, the entanglement entropy across the bipartition is 
\begin{equation}
    S(\mathcal{H}_1 : \mathcal{H}_2) = -\sum_i \lambda_i^2 \ln \lambda_i^2,
    \label{Eq:EntanglementEntropy}
\end{equation}
which coincides with the von Neumann entropy of either reduced density matrix. In this work, we systematically investigate the entanglement entropy generated across various bipartitions of the $N$-qubit system in the context of solving the MaxCut problem using QAOA and AQC.

\section{Methods}
\label{Section:Methods}

\subsection{QAOA Training Protocol}
\label{Section:OptimizationProcedure}
In this section, we describe the classical training procedure employed to determine the optimized variational parameters. For convenience, we collectively denote by $\phi^p_j$ the initial values of the variational parameters for a $p$-QAOA, where $\phi^p_j$ simultaneously represents both $\beta_j$ and $\gamma_j$. Similarly, we denote by $(\phi^p_j)^*$ the components of the optimized variational parameters, encompassing both $(\beta^p_j)^*$ and $(\gamma^p_j)^*$.

In the high-dimensional variational parameter space, a randomly chosen initial guess is likely to lead the classical optimizer toward a local minimum and a barren plateau \cite{Larocca_2025_BarrenPlateaus,Sack_QAInitialization_2021}. Consequently, it is essential to provide the optimizer with an informed initial guess to quickly drive the variational parameters towards the global minimum of $F_p(\vec{\boldsymbol{\beta}}_p, \vec{\boldsymbol{\gamma}}_p)$. Numerous studies have demonstrated that the optimized QAOA variational parameters for different instances of the MaxCut problem exhibit patterns \cite{Zhou2020PRX,Akshay2021ParameterConcentrationsQAOA,He2024ParameterSettingQAOA}; for a review, see Ref.~\cite{Blekos2024Review}. Motivated by these findings, heuristic strategies for generating informed initial guesses for the variational parameters have been proposed.

In Ref.~\cite{Zhou2020PRX}, it is observed that the optimized variational parameters $(\phi^p_j)^*$ vary smoothly as a function of $1 \leq j \leq p$. Based on this observation, a heuristic strategy termed \texttt{INTERP} was introduced, wherein an initial guess for the variational parameters of a $p$-QAOA circuit is generated by the optimized parameters of a $(p-1)$-QAOA circuit. The protocol for the initial guess, adjusted to the notation of this section, is given by
\begin{equation*}
    \phi^{p}_j = \frac{j-1}{p-1} \left( \phi^{p-1}_{j-1} \right)^{*} + \frac{p-j}{p-1} \left( \phi^{p-1}_j \right)^{*},
\end{equation*}
with $1 \leq j \leq p$. 

Additionally, Ref.~\cite{Lee2023InitializationStrategy} investigates the behavior of the optimized angles $(\phi^p_j)^*$, for a fixed $j$, as the number of QAOA layers $p$ increases. Based on these observations, a bilinear initialization strategy was proposed, wherein an initial guess for the variational parameters of a $p$-QAOA circuit is generated by exploiting approximate linear trends in both the angle index and the circuit depth. The bilinear initialization strategy is summarized as follows:
\begin{eqnarray*}
    \phi^p_j &=& 2(\phi^{p-1}_j)^* - (\phi^{p-2}_j)^*, \quad j \leq p-2, \nonumber \\ 
    \phi^p_j &=& (\phi^{p-1}_j)^* + (\phi^{p-1}_{j-1})^* - (\phi^{p-2}_{j-1})^*, \quad j = p-1, \\
    \phi^p_j &=& 2\phi^p_{j-1} - \phi^p_{j-2}, \quad j = p. \nonumber
\end{eqnarray*}

In this work, we implement a training strategy that combines the \texttt{INTERP} and bilinear initialization protocols. Following Ref.~\cite{Lee2023InitializationStrategy}, we first perform an exhaustive search over the variational parameters for circuit depths $p =1,2$. For $p=1$, the cost function is evaluated at 100 points on a $10 \times 10$ grid, and the optimization is performed starting from the three initial guesses with the lowest cost. The additional optimization is performed to ensure that the global minimum of $F_p(\vec{\boldsymbol{\beta}}_p, \vec{\boldsymbol{\gamma}}_p)$ is reached. For $p=2$, the optimization is initialized using the parameters optimized at $p=1$, combined with a grid of 100 points for the two additional angles. All angles are then optimized, and the procedure is repeated for the three initializations with the lowest cost. 

For $p \geq 3$, we propose a new initialization scheme in which the optimization is initialized using either the bilinear or interpolation strategy, depending on which yields a lower initial cost function value. As shown in Appendix \ref{Section:INTERP_Bilinear_appendix}, the protocol that produces the best initial guess depends on the specific problem instance, which motivates the need to incorporate both initialization protocols. Due to the $\mathbb{Z}_2$ symmetry of the Hamiltonian in Eq.~\eqref{Eq:HamiltonianMaxCutDegenerate}, the variational parameters $\beta_j$ can be restricted to the interval $ \left[-\frac{\pi}{4}, \frac{\pi}{4}\right)$
for all $j = 1, 2, \dots, p$. This symmetry has been identified in previous works~\cite{Zhou2020PRX}. Also, we restrict $\gamma_j$ to the interval $[0, 2\pi)$. Optimization is carried out using the L-BFGS-B algorithm with bound constraints, as implemented in the \texttt{SciPy} library. In our implementation, we set the maximum number of iterations \texttt{maxiter} = 3000 and the function tolerance (i.e., the termination criterion) \texttt{ftol} = $10^{-6}$. Finally, the quantum circuits are simulated using state vector simulation.

\subsection{Entanglement entropy calculation}

Utilizing the optimized variational parameters $(\phi^p_j)^*$ for a $p$-QAOA, we evaluate the von Neumann entanglement entropy after each layer $\ell \leq p$. Specifically, the entropy is computed for the state $\ket{\psi_\ell(\vec{\boldsymbol{\beta}}_\ell, \vec{\boldsymbol{\gamma}}_\ell)}$ described in Eq.~\eqref{Eq:VariationalStatevector}, where $\vec{\boldsymbol{\beta}}_\ell=\left( (\beta^p _1)^*,(\beta^p_2)^*,\dots,(\beta^p _{\ell})^*\right)$ and $\vec{\boldsymbol{\gamma}}_\ell=\left((\gamma^p _1)^*,(\gamma^p _2)^*,\dots,(\gamma^p _{\ell})^*\right)$. We define the entanglement at a specific layer $\ell$ and circuit depth $p$, $S^p_{\ell}(N_1; N) = \langle S^p_{\ell}(\mathcal{H}_1 : \mathcal{H}_2) \rangle$, as the average von Neumann entropy over an ensemble of bipartitions of an $N$-qubit system into sets of sizes $N_1$ and $N_2$. We define the bipartition with $N_1=N_2=N/2$ as the \emph{balanced} bipartition, and refer to any other bipartition with $N_1+N_2=N$ as an \emph{unbalanced} bipartition. Since the total number of qubits is fixed, it is sufficient to specify $N_1$, with $N_2 = N - N_1$. From this entropy measure, the maximum entropy across all layers of the $p$-QAOA is
\begin{equation}
    \label{Eq:SpDefinition}
    S^p(N_1; N) = \max_{\ell \leq p} S^p_{\ell}(N_1; N).
\end{equation}
The layerwise entropy $S^p_{\ell}(N_1; N)$ contrasts with $S^p(N_1; N)$ in that the latter identifies the layer within a fixed depth $p$ where entanglement reaches its maximum. 

We complement these quantities with the maximum entropy for a given $p_{\max}$ budget. Thus, given a maximum QAOA depth, this metric quantifies the entropy required to solve the computational problem across any available circuit depth. We refer to this quantity as the required entanglement entropy for QAOA and denote it as
\begin{equation}
    S^{\mathrm{req}}(N_1; N) = \max_{p \leq p_{\max}} S^p(N_1; N),
    \label{Eq:required_ent_definition}
\end{equation}
where $p_{\max} = 10 $ is the maximum QAOA depth considered in this work. To facilitate the presentation of the numerical analysis in the following sections, we also define the normalized quantity
\begin{equation}
    \tilde{S}^{\mathrm{req}}(N_1;N) = \frac{S^{\mathrm{req}}(N_1; N)}{S^{\mathrm{req}}(N/2; N)}, 
    \label{Eq:normalized_Sreq_numerics}
\end{equation}
where the normalization is given by the numerical data for the balanced bipartition.

In the subsequent analysis, we estimate $S^p_{\ell}(N_1; N)$ (and therefore obtain $S^p(N_1; N)$, and $S^{\text{req}}(N_1; N)$) for each graph by sampling an ensemble consisting of 5\% of all possible bipartitions. For each graph, the estimated value corresponds to the average of the ensemble while the error corresponds to the standard deviation. For the average behavior across all graphs, we report the mean value together with the standard error of the mean, obtained through error propagation from the uncertainties of the individual graphs. Further details that validate this sampling strategy are provided in Appendix~\ref{Section:SamplingStrategyAppendix}.

\subsection{Problem Specifications}

We investigate how entanglement in QAOA scales with system size $N$ using three datasets. In all datasets, every edge connects a distinct pair of vertices and is assigned a weight drawn from the uniform distribution $U(0,1)$. The system sizes tested range from $ 8 \leq N \leq 16 $.

\textbf{Dataset KR}: This dataset is used to quantify the entanglement requirements for a fixed class of graphs. It includes $k$-regular graphs with $2 \leq k \leq 7$, and provides a natural setting for studying the interplay between entanglement and per-qubit connectivity, $k$. 
For each pair $(N, k)$, we generate 150 distinct $k$-regular graphs.

\textbf{Dataset CW:} This dataset consists of 150 weighted complete graphs for each value of $N$. We distinguish this class of graphs from Dataset KR because, in this case, the connectivity per qubit scales with system size. In particular, each qubit is connected to all other $N-1$ qubits.

\textbf{Dataset ED}: This dataset is used to investigate the entanglement requirements of arbitrary graphs with $N$ qubits and $|E|$ edges, without restrictions on the graph type. To ensure that the problem is mapped to an $N$-qubit Hamiltonian (and not to a smaller system), we require that the minimum degree of each qubit is one, which sets the minimum number of edges to $|E| = \lceil \frac{N}{2} \rceil$; $\lceil x \rceil$ is the ceiling function. The maximum number of edges is $|E|=\frac{N(N-1)}{2}$, corresponding to complete graphs. Based on this range, we define the edge density
\begin{equation}
    \alpha = \frac{2|E|}{N(N-1)},
\end{equation}
which will be used to characterize the graphs in the analyses below. For each pair ($N,|E|$), 150 different graphs are generated.

\section{Impact of Training Protocol on the Entanglement Profile}
\label{Section:optimization_strategy_effect_qaoa}

Here, we show that there is interplay between the training protocol and the observed entanglement profile of QAOA. Namely, if the classical optimizer does not lead to sufficiently high approximation ratios, the true underlying structure of the entanglement dynamics can be concealed. This underscores the importance of leveraging a high-performing training routine when performing an analysis of QAOA’s entanglement dynamics.

In order to highlight this observation, we compare the training routine discussed in Sec.~\ref{Section:OptimizationProcedure}, with related training strategies~\cite{DupontEntanglementPerspective2022,Usman2024CalibratingRoleOfEntanglement}. The key difference between the two arises from distinct initialization strategy—the former relies on informed initial guesses, while the latter utilizes random variational parameters. As we will show, this subtle difference can result in a significant impact on the entanglement profile of QAOA. Importantly, these characteristics can be dependent upon the structure of the graph.

\subsection{Weighted Complete Graphs}
\label{Section:Optimizer_Entanglement_Complete_Graphs}
\begin{figure}[t!]
    \centering
    \includegraphics[width=\columnwidth,]{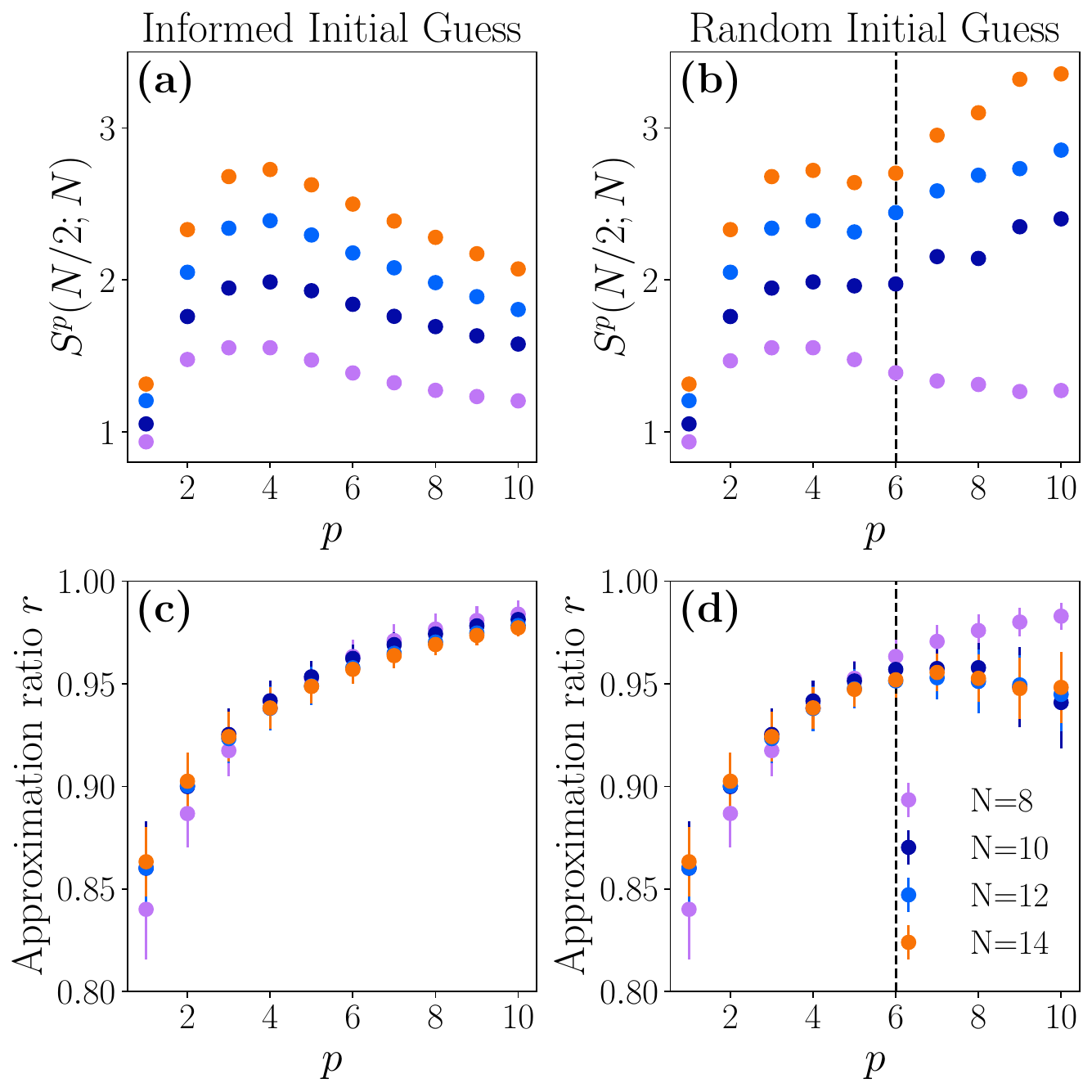}
    \caption{Comparison of random and informed initialization strategies. Panels (a) and (b) present the average  $S^p(N/2; N)$ as a function of QAOA depth $p$ for informed and random initialization, respectively. Panels (c) and (d) show the corresponding approximation ratios. Results are averaged over 50 random instances of weighted complete graphs.  Error bars indicate the standard error of the mean and are too small to be visible in panels (a) and (b). The performance of the classical optimizer can affect the entanglement profile of complete graphs.}
    \label{Fig:entanglement_optimizer_AppendixB}
\end{figure}

We begin by investigating the impact of the training strategy on the entanglement profile of Dataset CW. MaxCut problems within this graph family commonly constitute hard optimization problems due to their high connectivity. From the QAOA perspective, this feature of the problem can place greater demands on the classical training protocol, requiring performant routines in order to identify viable solutions. This implies that Dataset CW is a suitable choice to investigate the relationship between training and the entanglement profile.

We implement two training strategies. The first, discussed in Sec.~\ref{Section:OptimizationProcedure}, leverages an informed initial guess that is performed once. In contrast, the second approach starts from randomly chosen variational parameters. This process is repeated $10^3$ times and only the parameters yielding the lowest cost are retained. Following the notation of Eq.~\eqref{Eq:VariationalStatevector}, the initial variational parameters for a $p$-QAOA are sampled uniformly: $\gamma_j \in [0, \pi)$ and $\beta_j \in [0, \frac{\pi}{2})$, $1 \leq j \leq p$. The BFGS optimizer is employed, and the optimization is unbounded ~\cite{DupontEntanglementPerspective2022}. A closely related method is followed in Ref.~\cite{Usman2024CalibratingRoleOfEntanglement}. 

The difference in initialization strategies is reflected in the entanglement profiles, shown in panels (a) and (b) of Fig.~\ref{Fig:entanglement_optimizer_AppendixB}.
The entanglement $S^p(N/2;N)$ reaches a maximum around the critical QAOA depth $p^* \approx 4$ and decreases at larger depths when employing our training strategy; see Fig.~\ref{Fig:entanglement_optimizer_AppendixB}(a). This result differs from the random initialization strategy, where a roughly monotonic increase in entanglement entropy is observed; see Fig.~\ref{Fig:entanglement_optimizer_AppendixB}(b). Specifically, while both methods yield similar entanglement profiles up to $p \approx 4$, the random-initialization strategy exhibits an increase in entanglement beyond $p=6$. 

The distinct entanglement profiles can be attributed to differences in the algorithmic performance. The results for the approximation ratio are presented in Fig.~\ref{Fig:entanglement_optimizer_AppendixB}(c) and (d). Both strategies exhibit similar performance for small values of $p$: the solution quality improves as $p$ increases up to approximately $p=6$. Beyond this point, however, the performance of the random-initialization strategy begins to deteriorate, while the performance of the informed-initialization strategy continues to improve. Notably, $p=6$ is also the QAOA depth at which the entanglement profiles start to diverge. It appears that, at this QAOA depth,  the parameter space becomes large enough that a randomly chosen initial point does not lie near the global minimum of the cost function. These results reveal that the choice in training protocol can play a critical role in the entanglement profile.

Furthermore, we note that the results of this section may provide insight into previous entanglement studies of QAOA. Specifically, Refs. \cite{Usman2024CalibratingRoleOfEntanglement,DupontCalibratingClassicalHardness2022} implement a random initialization protocol---similar to the one used in this section---to study the connections between QAOA entanglement and the so-called simulation fidelity (i.e., the cost of simulating QAOA via MPSs). In the case of unweighted $k$-regular graphs, Ref.~\cite{Usman2024CalibratingRoleOfEntanglement,DupontCalibratingClassicalHardness2022} identify a scaling law for the simulation fidelity that relates the entanglement per qubit to the bond dimension required by the MPS, achieving a fidelity of unity when the cost of simulation is the same as a simulation with bond dimension $2^{N/2}$, where $N$ is the number of qubits. The scaling law holds only for shallow QAOA depths before the entanglement entropy reaches a maximum. However, for CW graphs, the scaling law persists for deeper QAOA circuits, which Ref.~\cite{Usman2024CalibratingRoleOfEntanglement} attributes to the absence of a peak in the entanglement. Based on our analysis, we conjecture that the latter analysis may suffer from insufficient classical training that conceals the true entanglement profile.

\subsection{$k$-Regular Graphs and Graphs of Varying Edge Density}
\label{Sec:impact_opt_QAOA_KR_ED_dataset}

\begin{figure*}[t!]
\centering
\includegraphics[width=0.9\textwidth]{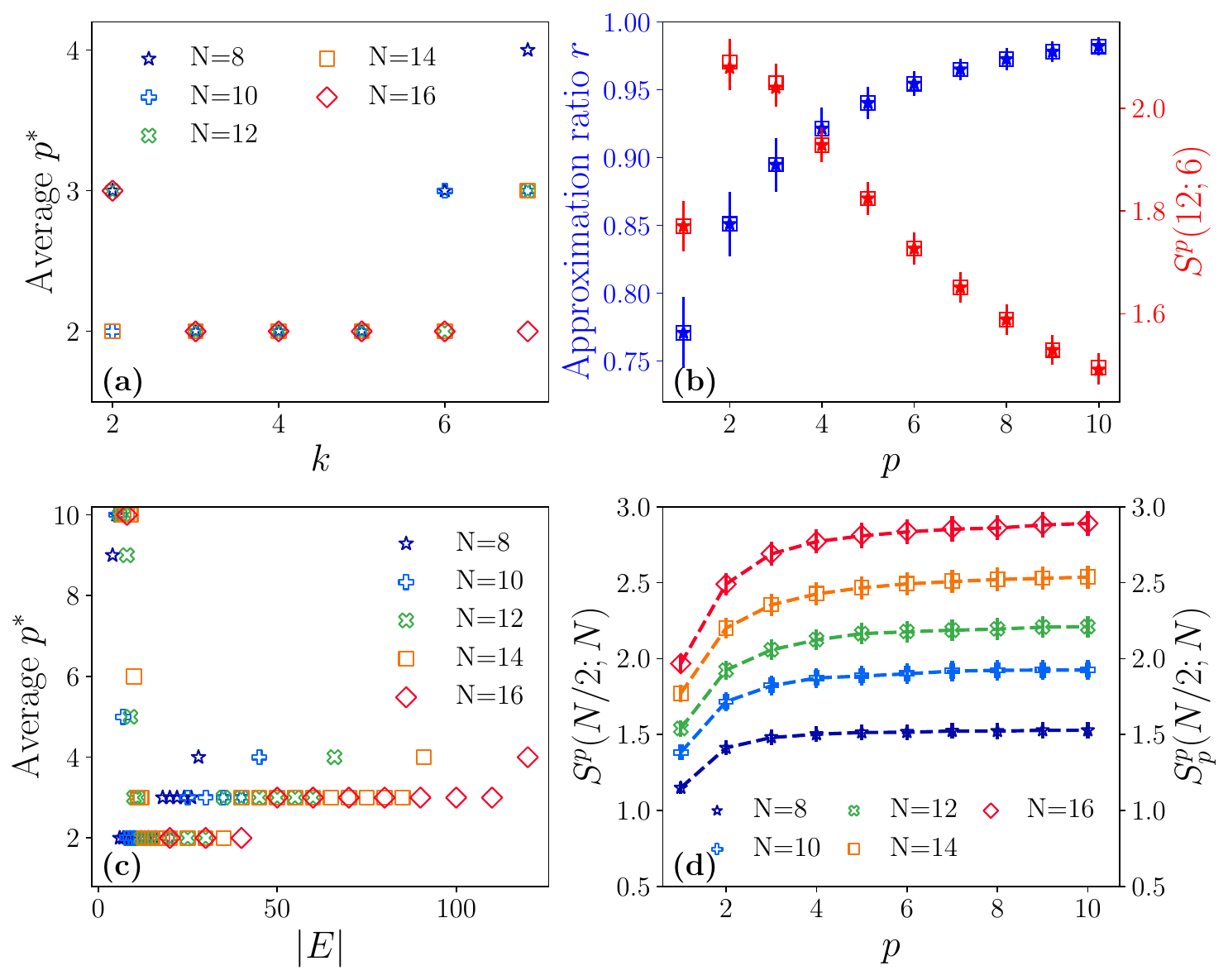}
\caption{Numerical simulations of $S^p(N/2;N)$ and $S^p_p(N/2;N)$ as a function of QAOA depth $p$ for Dataset KR and Dataset ED. The average number of QAOA layers $p^{*}$ required to reach $S^{p^*}(N/2; N) = S^{\mathrm{req}}(N/2; N)$, for Dataset KR and Dataset ED are shown in panels (a) and (c), respectively, for different system sizes $N$.  
In addition, panel (b) shows simulations for weighted 3-regular graphs with $N=12$ using informed (stars) and random initial guesses (squares) for optimization. The entanglement entropy $S_{p}^{p}(N/2; N)$ at the final layer of the QAOA circuit (dashed line), and the maximum entanglement entropy $ S^{p}(N/2; N)$ (data points with different markers) as a function of QAOA depth $p$ are shown for sparse graphs with $|E| = N/2$ in panel (d). Error bars denote the standard error of the mean and are smaller than the marker size. For both Dataset KR and Dataset ED, $p^*$ typically requires shallow QAOA circuits, except in the low edge density regime where the number of solution bitstrings is large.
}
\label{Fig:peak_QAOA_Depth}
\end{figure*}

In the previous subsection, we demonstrated that the choice of classical training protocol affects the entanglement profile of complete graphs. Here, we apply our training method to Datasets KR and ED to investigate whether this observation extends to other graph instances.

We begin by discussing results for the weighted $k$-regular graphs (i.e., Dataset KR). In Fig.~\ref{Fig:peak_QAOA_Depth}(a), the QAOA depth at which $S^p(N/2;N)$ reaches its maximum is compared against $k$. We observe that a peak in the entanglement typically occurs for $p^* \leq 4$. Notably, this peak occurs at relatively shallow circuit depths, implying that QAOA does not need many layers to achieve its maximum required entanglement. These findings are consistent with previous studies of 3-regular and 4-regular graphs~\cite{Usman2024CalibratingRoleOfEntanglement}, where the random initialization strategy is implemented. As such, the entanglement profile of $k$-regular graphs does not appear to exhibit strong dependence on the training routine. 

Similarly, both training strategies exhibit the same trends in the approximation ratio. In Fig.~\ref{Fig:peak_QAOA_Depth}(b), we specifically highlight this behavior for weighted 3-regular graphs and a problem size of $N=12$. Both the approximation ratio and entanglement profile are shown as a function of $p$ for the informed initialization (stars) and random initialization (squares). In contrast to the case of complete graphs, both training protocols show a monotonically increasing approximation ratio, which is reflected in the presence of a peak in the entanglement for both cases. Alternative $k$ and $N$ values convey equivalent characteristics.

Similar results are observed for Dataset ED. For graphs with a large number of edges $|E|$, the general trend resembles that of $k$-regular graphs, where the maximum entanglement is obtained for $p^* \leq 4$; see Fig.~\ref{Fig:peak_QAOA_Depth}(c). The  presence of a peak is in agreement with the entanglement profile of grid graphs---graphs with varying edge density, but defined structure~\cite{Usman2024CalibratingRoleOfEntanglement}—--indicating that the training strategy does not alter the entanglement profile of graphs in Dataset ED.

An exception to the trend is observed in the low edge density regime. Specifically, for $|E| = N/2$, it is found that $p^*$ approaches  $p^*=10$. To examine this behavior more closely, we plot $S^p(N/2; N)$ and $S^p_p(N/2; N)$ as a function of QAOA depth in Fig.~\ref{Fig:peak_QAOA_Depth}(d). The entanglement entropy does not exhibit a clear peak within the accessible $p$. This trend can be understood by considering the dimension of the solution space in these sparse graphs. Here, the graph decomposes into $N/2$ disconnected two-qubit subgraphs, each admitting two solutions. As a result, the total number of solutions is $2^{N/2}$. The QAOA circuit must therefore generate entanglement across a large superposition of solution states, requiring a greater number of layers to do so. As such, the entanglement profile as a function of QAOA order is described by an increase in entanglement with $p$, rather than an entanglement barrier at a critical $p$. 

As the edge density increases, the solution space becomes more constrained. In the absence of additional degeneracies, the number of solutions approaches, on average, two. Thus, ideally resulting in an entanglement barrier at an intermediate QAOA depth and low entanglement at the end of the QAOA evolution. We note that this behavior is also observed for complete graphs. This analysis highlights how highly degenerate solution spaces, as found in sparse graphs, can significantly affect entanglement growth. In Sec.~\ref{Section:balanced_bipartition_dataset1_dataset2_QAOA}, we further show its effect on the entanglement scaling with system size.

Overall, the presence of a peak in the entanglement profile appears to be independent of the training routine for Datasets KR and ED. This observation is in contrast to the analysis of Sec.~\ref{Section:Optimizer_Entanglement_Complete_Graphs}. We attribute this difference to the extensive connectivity of complete graphs, which makes them harder problems to solve. 

\section{Problem Structure and QAOA Entanglement Scaling}
\label{Section:problem_structure_ent_scaling_QAOA}

\subsection{Entanglement Scaling for $k$-Regular and Complete Graphs}
\label{Section:balanced_bipartition_dataset1_QAOA}

\begin{figure}[t!]
\centering
\includegraphics[width=0.48\textwidth,keepaspectratio]{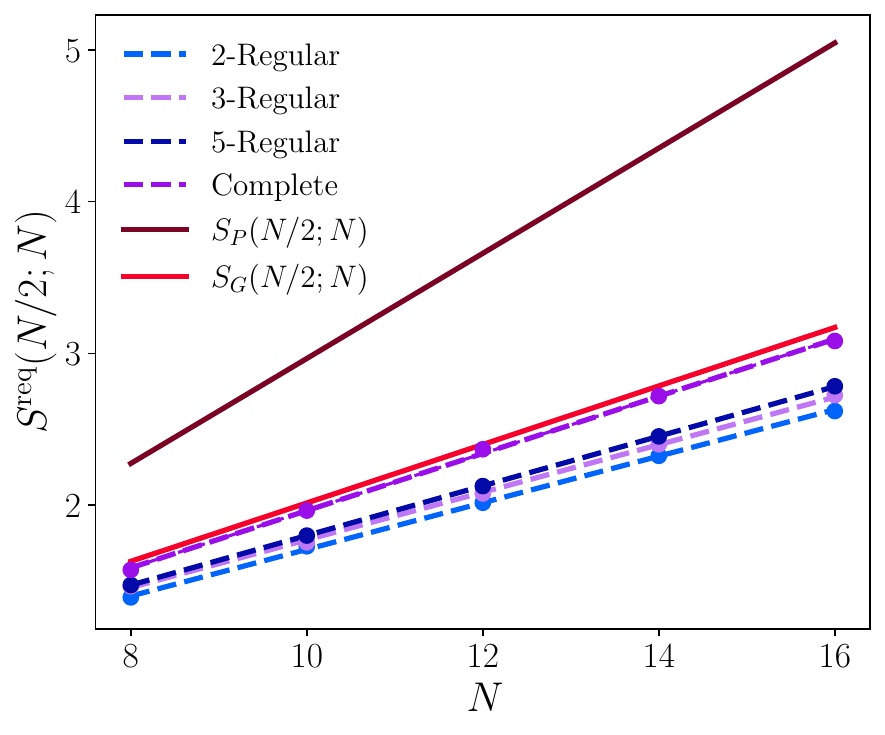}
\caption{Scaling of $S^{\mathrm{req}}(N/2;N)$ in QAOA for Datasets KR and CW. $S_P(N/2; N)$ and $S_G(N/2; N)$ denote analytical predictions for Haar random states (i.e. Page Value) and fermionic Gaussian states, respectively; see Sec.~\ref{subsec:balanced-fgs} for further details.
Error bars denote the standard error of the mean and are smaller than the marker size. The required entanglement is upper-bounded by that of fermionic Gaussian states and scales linearly with system size for both training schemes implemented in this work.}
\label{Fig:EntanglementScalingHalf}
\end{figure}

In the previous section, we investigated the number of QAOA layers required to achieve the maximum entanglement entropy and its relation to problem structure and training strategy. Here, we examine the scaling of this maximum, i.e., $S^{\text{req}}(N/2; N)$, as a function of system size. We assess whether the scaling previously observed for the MaxCut problem persists as we change the structure of the underlying graph. 

To this end, we consider a volume-law scaling described by
\begin{equation}
    F(N) = \gamma^{\mathrm{QAOA}} N + \delta^{\mathrm{QAOA}},
    \label{eq:fit}
\end{equation}
where $\gamma^{\mathrm{QAOA}}$ and $\delta^{\mathrm{QAOA}}$ are fit parameters. Similar fits were previously used to describe MaxCut unweighted 3-regular and weighted complete graphs~\cite{DupontEntanglementPerspective2022}. Thus, they provide a reasonable assumption for the MaxCut problems considered in this work.

We assess the volume-law entanglement scaling using the informed initial-guess optimizer introduced in this work. The scaling behavior is initially investigated numerically for a broader set of graph families in Dataset KR, including $k$-regular graphs with $k=2,3,5$, and Dataset CW. The results for these graph families and different system sizes are presented in Fig.~\ref{Fig:EntanglementScalingHalf}. The numerical data remain well described by Eq.~\eqref{eq:fit}, with the slopes given by $\gamma^{\mathrm{QAOA}}_{2R} \approx 0.153$, $\gamma^{\mathrm{QAOA}}_{3R} \approx 0.157$, $\gamma^{\mathrm{QAOA}}_{5R} \approx 0.163$, and $\gamma^{\mathrm{QAOA}}_{CW} \approx 0.188$ for 2-regular, 3-regular, 5-regular, and weighted complete graphs, respectively. This analysis confirms that the volume-law behavior persists across different graph families, despite variations in the classical training strategies.

The scaling comparison between graph types reveals intuitive behavior. The maximum entanglement tends to increase more rapidly as $k$ increases. Weighted complete graphs afford the largest growth in entanglement, thus agreeing with the expectation that problems with higher connectivity per qubit require a larger amount of entanglement.

\subsection{Entanglement Scaling with Graph Edge Density}
\label{Section:balanced_bipartition_dataset1_dataset2_QAOA}
\begin{figure*}[t!]
    \centering
    \includegraphics[width=0.9\textwidth]{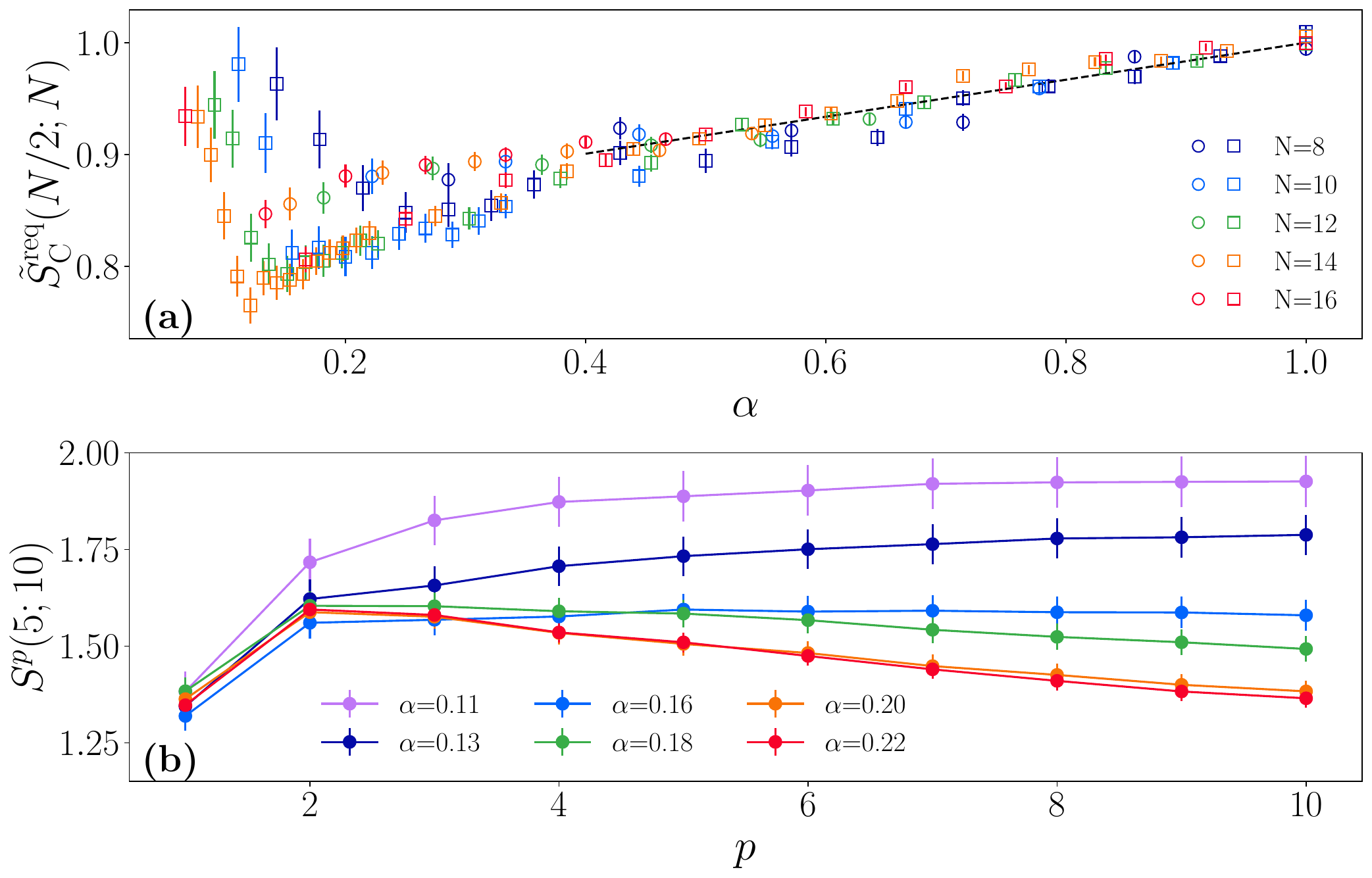}
    \caption{Numerical simulations for entanglement requirement of QAOA for arbitrary graphs. Results from Dataset KR (circles) and Datasets CW and ED (squares) are shown. Panel (a) shows results, normalized by complete graphs, as a function of edge density. Dashed lines indicate the linear fit described in the text. Panel (b) shows $S^p(N/2;N)$ versus QAOA depth for $N=10$ qubits with different edge counts. Error bars denote the standard error of the mean. The fit accurately captures the entanglement in the high edge-density regime.}
    \label{Fig:MaxEntGeneralGraphQAOA}
\end{figure*}
 
Although the constraints imposed by Datasets KR and CW facilitate controlled comparisons, they inherently restrict the diversity of graphs examined. These constraints limit the generality of conclusions regarding entanglement scaling. In this section, we relax these connectivity restrictions by incorporating results from Dataset ED, which consists of random graphs characterized by varying numbers of edges. This dataset enables a more generalized investigation into the relationship between the graph's features and the entanglement generated by QAOA.

We investigate the universality of the volume scaling by examining the required entanglement for all three datasets in Fig.~\ref{Fig:MaxEntGeneralGraphQAOA}(a). Dataset KR is denoted by open circles, where each dataset has been expressed in terms of the edge density $\alpha$. Datasets ED and CW are displayed by open squares. The required entanglement is normalized by the fitted scaling law for CW, $F_C(N)$, as this represents a bound on the entanglement

\begin{equation}
\tilde{S}^{\mathrm{req}}_{\mathrm{C}}(N_1;N) = \frac{S^{\mathrm{req}}(N_1; N)}{F_C(N)}.
\label{Eq:normalized_Sreq_complete_fit}
\end{equation}
Furthermore, the ratio illustrates a commonality in entanglement scalings among the three datasets. Each is subject to a period of declining entanglement at low edge density before reaching a minimum at a critical density $\alpha_c$. Thereafter, the required entanglement steadily increases with increasing edge density.

We attribute the reduction in entanglement at low edge density to the presence of MaxCut problems with highly degenerate ground state manifolds. This behavior was previously observed in Sec.~\ref{Sec:impact_opt_QAOA_KR_ED_dataset} when investigating the QAOA order where maximum entanglement is achieved, and it is further illustrated in Fig.~\ref{Fig:MaxEntGeneralGraphQAOA}(b). As $\alpha$ approaches $\alpha_c$, the entanglement steadily declines due to a reduction in the ground state degeneracy. The barrier emerges for $\alpha\geq \alpha_c$, leading to a fundamentally distinct scaling in the required entanglement.

Beyond the critical edge density the required entanglement scales relatively consistently for all datasets. We observe a trend that can be captured by the scaling law 
\begin{equation}
    \tilde{F}_C(\alpha)= m^\mathrm{QAOA} (\alpha - 1) + 1,
    \label{Eq:MaxEntGeneralGraph}
\end{equation}
where $m^{\mathrm{QAOA}} \approx 0.165$. Note that for $\alpha = 1$, this expression recovers the volume law of complete graphs. Furthermore, note that this expression does not explicitly convey dependence on $k$, differing from the observed behavior in Sec.~\ref{Section:balanced_bipartition_dataset1_QAOA}. 
Nevertheless, as illustrated in Fig.~\ref{Fig:MaxEntGeneralGraphQAOA}(a), Eq.~\eqref{Eq:MaxEntGeneralGraph} provides a reasonable approximation to the required entanglement for all datasets, including Dataset KR, and for all system sizes considered.

\subsection{QAOA Entanglement Scaling Bounds: Relations to Fermionic Gaussian States}
\label{subsec:balanced-fgs}
Previous assessments of QAOA entanglement scaling have sought to develop an upper bound. Since analytical techniques tend to be difficult to leverage in the setting of QAOA, numerical approaches have been preferable. Comparisons have aimed to utilize the average entanglement generated by Haar-random states (i.e., the Page Value~\cite{PageValueRandom1993}) to upper bound entanglement for the MaxCut problem based on QAOA dynamical simulations of an ensemble of problem instances. The justification relies on demonstrating that the entanglement spectrum of QAOA follows the Marchenko–Pastur distribution~\cite{DupontEntanglementPerspective2022}, implying that the circuit exhibits sufficient randomness to yield volume-law entanglement scaling. 

In Fig.~\ref{Fig:EntanglementScalingHalf}, we compare the Page value to the scaling laws obtained for Dataset KR. In particular, we compare our numerical fits to 
\begin{equation}
    S_P(N_1; N) = \Psi(2^N + 1) - \Psi(2^{N - N_1} + 1) - \frac{2^{N_1} - 1}{2^{N - N_1 + 1}},
    \label{Eq:PageRandom}
\end{equation}
where $\Psi(z) = \Gamma'(z)/\Gamma(z)$ is the digamma function~\cite{PageValueRandomProofFoon1994,PageValueRandomProofSen1996,PageValueRandomProofRuiz1995}. Fig.~\ref{Fig:EntanglementScalingHalf} is focused on balanced bipartitions and thus, $N_1 = N/2$. While upper bounding the required entanglement, the Page Value does not provide a tight bound. Drawing on existing work in the study of random states, we aim to identify a tighter bound on the entanglement generated by QAOA.  

To this end, we examine the potential relationship between QAOA and random fermionic Gaussian states ~\cite{Bianchi2021FermionicGaussianStates}. FGSs form a class of quantum states generated by Hamiltonians that are quadratic in the fermionic operators. They are fully and efficiently characterized by a correlation matrix containing all two-point correlation functions~\cite{Surace_FGS_notes_2022}. FGSs have been shown to approximate both the entanglement entropy and the ground-state energy of transverse-field Ising models ~\cite{Kaicher_Mean_Field_2023}, a variation of the MaxCut Hamiltonian given in Eq.~\eqref{Eq:HamiltonianMaxCutDegenerate}. In addition, it has been shown that the efficiency of the state preparation in QAOA is correlated with the distance between the target state and the manifold of FGSs~\cite{Matos_State_Preparation_FGS_2021}. This is further corroborated from a symmetry perspective in Ref.~\cite{Matos_VQAs_free_fermions_2023}. 

Here, we examine the connection between FGS and QAOA via the required entanglement scaling. We consider an expression for the average entanglement entropy of random FGSs given by 
\begin{eqnarray}
\label{Eq:PageFermionicGaussianStates}
\begin{aligned}
   &S_{G}(N_1; N) = \left( \frac{1}{2} + N_1 - N \right) \Psi(2N - 2N_1) - N_1 \\
   &+ \left( N - \frac{1}{2} \right) \Psi(2N) + \left( \frac{1}{4} - N_1 \right) \Psi(N) - \frac{1}{4} \Psi(N - N_1),
\end{aligned}
\end{eqnarray}
where $N_1 \leq N - N_1$ is imposed~\cite{Bianchi2021FermionicGaussianStates}. In the large system-size limit and considering a balanced bipartition, the expression reduces to 
\begin{equation*}
    S_G\left(N_1 = \frac{N}{2}; N\right) \overset{N \rightarrow \infty}{\approx} N\left(\ln 2 - \tfrac{1}{2}\right) + \tfrac{1}{4}(1 - \ln 2).
\end{equation*}
The former is displayed in Fig.~\ref{Fig:EntanglementScalingHalf} for comparison against the numerical fits. Note that this expression provides a tight upper bound to the estimated scaling obtained for $k$-regular and weighted complete graphs. In particular, for Dataset CW, we find that it only differs by approximately $2.7\%$.

This result has several implications. First, signatures of FGSs in QAOA states suggests that FGSs may have utility as high-quality initial guesses that accelerate algorithmic convergence. Preparing such states is feasible using matchgate circuits \cite{Bejan_FGS_Matchgate_2025}. In addition, a connection with FGSs can be leveraged to develop more efficient QAOA ans\"atze. For instance, Ref.~\cite{Dupont_RelaxAndRoundQAOA_2024} proposes a QAOA variant which exploits the qubit correlation matrix to enhance algorithmic performance. FGSs are uniquely characterized by the two-point correlation matrix, and therefore offer a new perspective for exploring such QAOA variants. Overall, while our results do not currently prove a connection between FGSs and QAOA, we believe that they do highlight a potentially interesting relationship that should be further explored. Connections between QAOA and such states may be key to providing analytical insight into performance guarantees and algorithmic hardness.

\subsection{Further Connections to Fermionic Gaussian States: Beyond Balanced Bipartitions}
\label{Section:entanglement_all_bipartitions_dataset1_QAOA}
\begin{figure}[t!]
\centering
\includegraphics[width=0.5\textwidth,keepaspectratio]{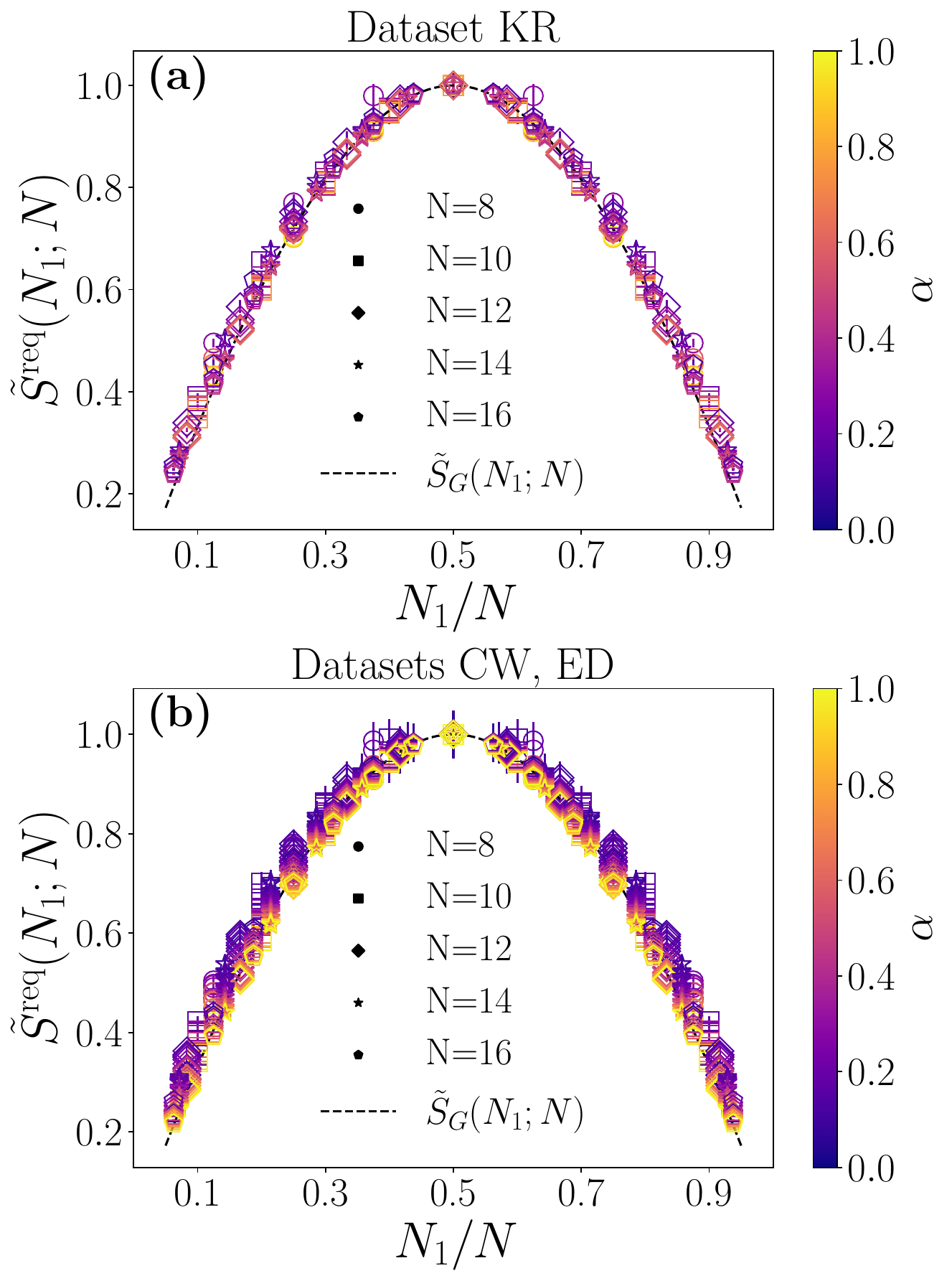}
\caption{Numerical simulations for the entanglement requirement in QAOA, across all bipartitions and different graph families. Data for (a) $k$-regular graphs from Dataset KR and (b) complete graphs and graphs with varying edges from Datasets CW, ED are shown. Different colors reflect the edge density, while different markers indicate the system size. Error bars denote the standard error of the mean. For graphs with large edge density, the numerical data collapse onto a single curve, which is well described by Eq.~\eqref{Eq:fermionic_Gaussian_normalized}.}
\label{Fig:CompareWithFermionicGaussianAll}
\end{figure}

Thus far, we have analyzed the scaling of entanglement for the balanced bipartition. In this subsection, we extend this analysis to unbalanced bipartitions, aiming to provide greater insight into the entanglement characteristics of QAOA. Studying the entanglement across various bipartitions can reveal critical features of entanglement growth of many-body systems, particularly in the context of entanglement dynamics following a quantum quench \cite{Nahum_Ent_Growth_Quench_2017}. Such quench dynamics bears a direct connection to the QAOA ansatz in Eq.~\eqref{Eq:VariationalStatevector}, which can be understood as a structured sequence of controlled quantum quenches. While we will not directly rely on previous quench dynamics results, we utilize this connection as motivation for studying unbalanced bipartitions for QAOA. As we show below, this analysis proves fruitful, indicating further connections between FGSs and QAOA entanglement growth.

We perform numerical simulations and investigate the dependence of the required entanglement on the subsystem size $N_1$. In Fig.~\ref{Fig:CompareWithFermionicGaussianAll}, the required entanglement as a function of $N_1/N$ is shown for Datasets KR and ED in panels (a) and (b), respectively. The required entanglement is normalized by the balanced partition entanglement $S^{\mathrm{req}}(N/2; N)$, according to Eq.~\eqref{Eq:normalized_Sreq_numerics}, as we anticipate this being the maximum over all $N_1$. The data has been colored by edge density $\alpha$ for all datasets.

The comparison indicates a clear structure for the required entanglement. As $N_1$ changes, the entanglement increases, reaching a maximum at $N_1=N/2$. Thereafter, the required entanglement decreases. The parabolic-like dependence is predominately invariant to $\alpha$, indicating that, for all datasets considered here, the unbalanced bipartition entanglement is not strongly influenced by edge density. 

In Sec.~\ref{subsec:balanced-fgs}, we provided empirical evidence for FGSs bounding the required entanglement for balanced bipartitions. Here, we examine whether the FGS connection can be extended to unbalanced bipartitions as well. To enable this comparison, we introduce a normalized version of Eq.~\eqref{Eq:PageFermionicGaussianStates}:
\begin{equation}
    \label{Eq:fermionic_Gaussian_normalized}
    \tilde{S}_G(N_1; N) = \frac{S_G(\min(N_1, N - N_1); N)}{S_G(N/2; N)}.
\end{equation}
Since Eq.~\eqref{Eq:PageFermionicGaussianStates} assumes $N_1 \leq N/2$, we swap $N_1 \leftrightarrow N_2$ when plotting the normalized version of Eq.~\eqref{Eq:fermionic_Gaussian_normalized} for $N_1 > N/2$ in Fig.~\ref{Fig:CompareWithFermionicGaussianAll}. The comparison reveals an agreement between the normalized QAOA entanglement requirements and the analytical expression of Eq.~\eqref{Eq:fermionic_Gaussian_normalized}. The coefficient of determination $R^2$, between the numerical data and Eq.~\eqref{Eq:fermionic_Gaussian_normalized} is $R^2 \geq 0.9$, improving to $R^2 \geq 0.99$ for large $\alpha$. This result suggests that the entanglement profile of QAOA, up to a normalization factor, is well described by the entanglement profile of FGSs. While conclusively identifying maximally entangled QAOA states as FGSs remains an open question, our findings uncover a relationship between the structure of different computational problem families and the presence of FGSs.

In the next sections, we turn our attention to AQC and its associated entanglement scaling. We focus specifically on the impact of the annealing protocol and problem structure on the entanglement profile.

\section{Impact of Annealing Schedule on AQC Entanglement Profile}
\label{Section:impact_annealing_schedule_AQC}

\begin{figure*}[t!]
    \centering
    \includegraphics[width=0.9\textwidth]{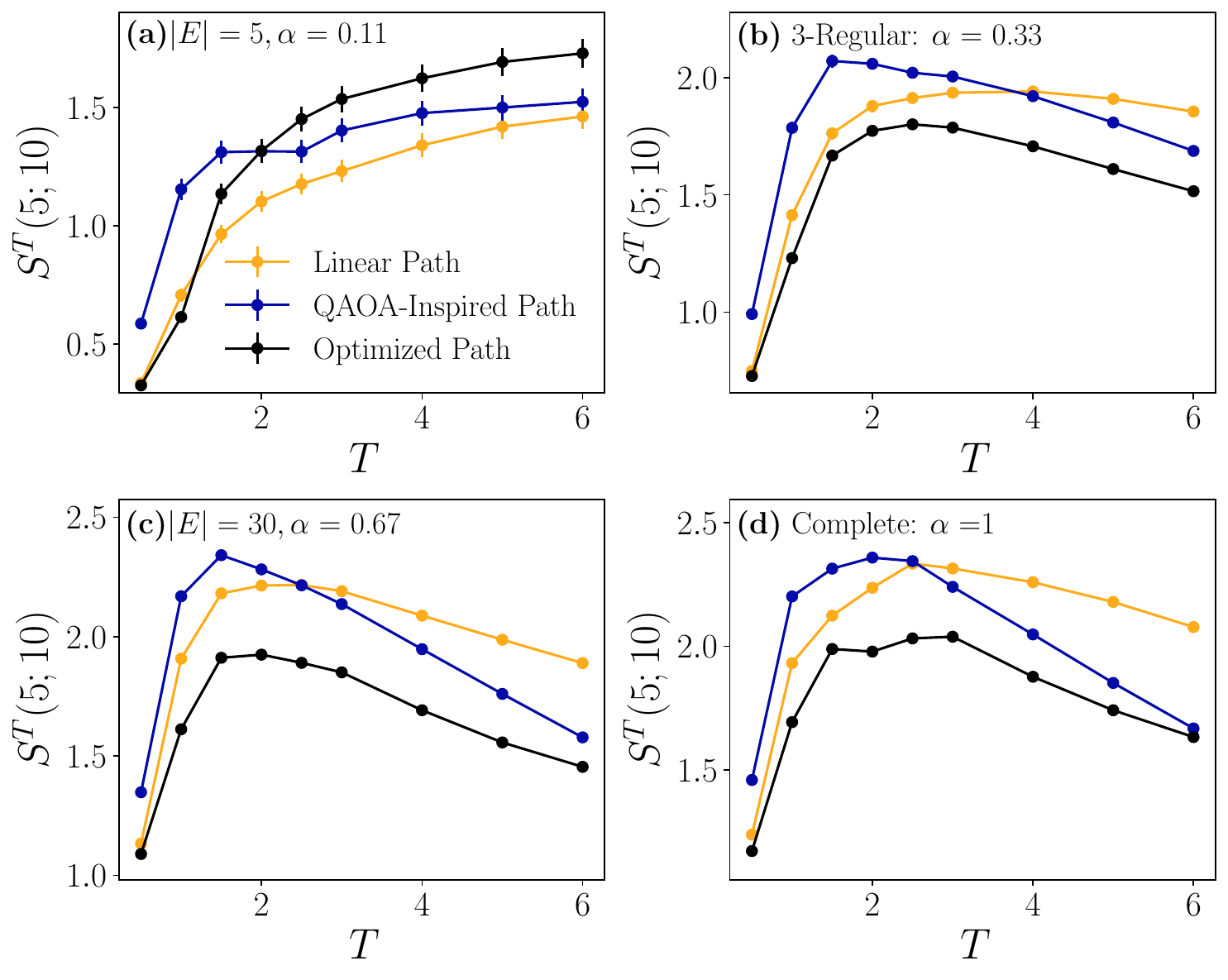}
    \caption{Numerical simulations of $S^T(N/2;N)$ for different annealing paths and graphs. Panels (a) and (c) correspond to instances from Dataset ED, panel (b) to Dataset KR, and panel (d) to Dataset CW. In all cases, the system size is $N=10$, and the balanced bipartition is considered. For each panel, the data are averaged over 150 problem instances. The optimized path generates more entanglement than the linear and QAOA-inspired paths at low-edge density. However, this trend reverses as the edge density increases, revealing an interplay between annealing path choice and graph features.}
    \label{Fig:Impact_Annealing_Path_AQC}
\end{figure*}

Here we investigate entanglement in AQC for the MaxCut problem. The goal of this section is to characterize the qualitative and quantitative features of the AQC entanglement profile, and examine its robustness with respect to the choice of the annealing path. This analysis is motivated by the results of Sec.~\ref{Section:optimization_strategy_effect_qaoa}, where we identified a dependence of the QAOA entanglement profile on the training protocol for complete graphs. As we show below, the entanglement profile in AQC is sensitive to the choice of the annealing schedule.

\subsection{Choice of annealing paths}
We consider three classes of annealing schedules. The first is the linear path $f(s)=s$, which has been previously used to explore entanglement in AQC for Dataset CW and unweighted 3-regular graphs~\cite{DupontEntanglementPerspective2022}. Here, we extend this analysis to Dataset ED to explore the interplay with graph features. 

Although the linear schedule is a paradigmatic choice, it is not optimal in terms of obtaining high-quality solutions. To explore the effect of schedule optimization, we introduce the following family of interpolation functions: 

\begin{equation*}
    f(s,k,c) = \frac{s^k}{s^k + (\frac{c}{1-c})^k (1-s)^k},
\end{equation*}
where $c,k$ are parameters to be optimized. This family covers a broad class of sigmoid-like functions that captures the linear path for $k=1$ and $c=1/2$ and paths that aim to slow near minimum energy gaps~\cite{roland2002aqc, rezakhani2009aqc, Quiroz_quantum_control_AQC_2019}. The optimized path is determined by selecting the parameters $c$ and $k$ that maximize the approximation ratio at the end of the evolution. The optimization is constrained to the parameter space $0<c<1$ and $0<k<2$.

Finally, we explore entanglement in AQC using translations from optimized QAOA circuits. We refer to this approach as the QAOA-inspired path. To this end, we implement the approach introduced in Ref.~\cite{Zhou2020PRX}, which defines a smooth path based on the optimized parameters of a $p$-QAOA. Since the $p$-QAOA ansatz in Eq.~\eqref{Eq:AdiabaticTheoremOriginal} is a discretization of the continuous evolution in AQC, we can interpret the sum of the variational parameters as the total annealing time $T_p = \sum^p_{j=1} (|(\beta^p_j)^*| + |(\gamma^p_j)^*|)$. A smooth annealing path can be constructed as follows \cite{Zhou2020PRX}:

\begin{eqnarray}
f (t_i) &=& \frac{(\gamma^p_i)^*}{|(\gamma^p_i)^*| + |(\beta^p_i)^*|}, \\ 
t_i &=& \sum^{i}_{j=1} (|(\gamma^p_j)^*| + |(\beta^p_j)^*|) - \frac{1}{2} (|(\gamma^p_i)^*|+|(\beta^p_i)^*|), \nonumber
\label{Eq:annealing_path_qaoa}
\end{eqnarray}
with the boundary conditions $f(t_i=0)=0$ and $f(t_i=T_p)=1$. For our analysis, $T_p$ is computed to define the annealing path as a function of the normalized time $f(s_{i,p})$, where $s_{i,p}=t_i/T_p$, and does not represent the total evolution time of the system $T$. To enable a consistent comparison, we consider the translated path derived from a $p$-QAOA with approximation ratio $r \geq 0.99$. We note that, achieving such a high approximation ratio requires extending the simulations beyond $p=10$, which was the maximum QAOA depth considered in Secs.~\ref{Section:optimization_strategy_effect_qaoa} and \ref{Section:problem_structure_ent_scaling_QAOA}.

For a given choice of annealing path, the time evolution over a total time $T$ is implemented using the fourth-order Suzuki–Trotter decomposition~\cite{Titum_phase_transition_quench_dynamics_2019,Susuzki_Approx_1990}, with a time step $\Delta t=0.1$. 

\subsection{Entanglement Dynamics and Choice of Annealing Path}
We start by investigating the interplay between entanglement dynamics in AQC and the choice of annealing path. Specifically, we examine how the maximum entanglement $S^T(N/2;N)$ varies with the evolution time $T$ as a function of the annealing schedule. The quantity $S^T(N/2;N)$ corresponds to the entanglement metric defined in Eq.~\eqref{Eq:SpDefinition} for QAOA, under the correspondences $T \leftrightarrow p$ and $t \leftrightarrow \ell$. 

The results for representative instances from Datasets CW, KR, and ED are presented in Fig.~\ref{Fig:Impact_Annealing_Path_AQC}. In the low edge-density regime shown in panel (a), entanglement increases as a function of $T$ for all interpolating paths considered. Similar to the discussion in Sec.~\ref {Sec:impact_opt_QAOA_KR_ED_dataset}, the absence of an entanglement peak can be attributed to the highly degenerate solution space. The same argument can be used to explain the relative entanglement generated across different methods. As shown in panel (a), the optimized path generates more entanglement than the linear and QAOA-inspired schedules. This can be understood as the effect of the schedule optimization: by achieving a higher approximation ratio at small times, the optimized schedule brings the system closer to the highly entangled ground state, thereby generating more entanglement. 

However, the entanglement dynamics of the different methods are sensitive to graph features. As the edge density increases (panels (b), (c), and (d) of Fig.~\ref{Fig:Impact_Annealing_Path_AQC}), the entanglement initially grows, reaches a peak, and subsequently decreases. In these regimes, the optimized path produces the least entanglement, while the QAOA-inspired path yields entanglement comparable to that of the linear path. From this analysis, we conclude that the AQC entanglement profile depends strongly on both the choice of the annealing schedule and the underlying graph features. Similar results are observed for other system sizes considered in this work.

\section{Problem structure and AQC Entanglement Scaling}
\label{Section:problem_structure_ent_scaling_AQC}

\subsection{Entanglement Scaling for $k$-Regular and Complete Graphs}
\begin{figure*}[t!]
    \centering
    \includegraphics[width=0.9\textwidth]{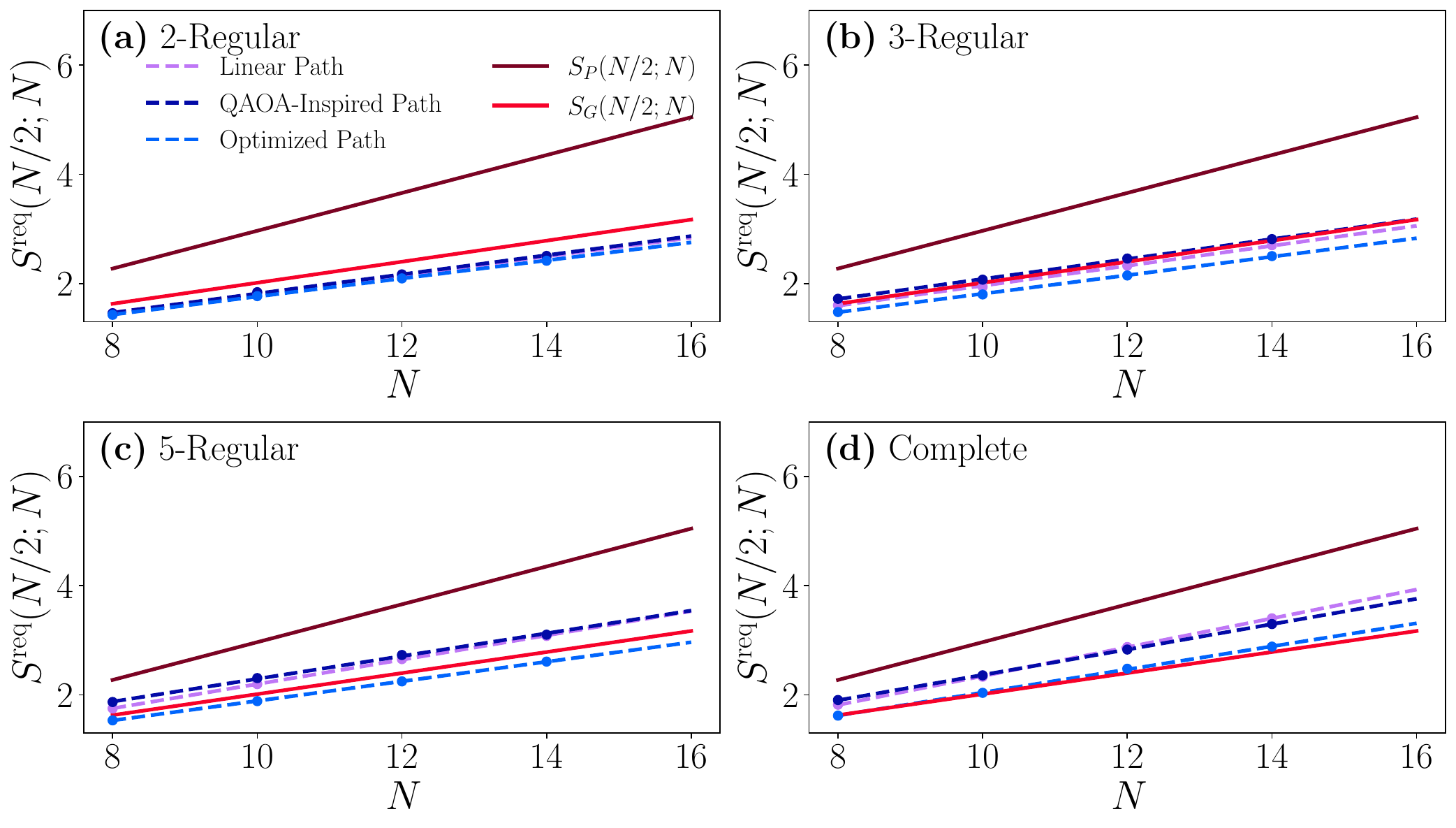}
    \caption{Scaling of $S^{\mathrm{req}}(N/2;N)$ in AQC for KR and CW datasets across multiple annealing schedules. The results are compared against the analytical benchmarks $S_P(N/2; N)$ and $S_G(N/2; N)$ of Eqs.~\eqref{Eq:PageRandom} and \eqref{Eq:PageFermionicGaussianStates}. Each data point represents an average over 150 problem instances. $S^{\mathrm{req}}(N/2;N)$ follows a volume law-scaling, with a slope that depends on the graph structure and the chosen annealing schedule.}  
    \label{Fig:volume_law_AQC}
\end{figure*}

In the previous section, we examined the features of the AQC entanglement profile and identified a peak value for most graph families, except those in the low-edge-density regime. Here, we investigate how this maximum scales with system size for different graph families in Datasets KR and CW. For AQC, the quantity of interest, $S^{req}(N/2;N)$, is defined according to Eq.~\eqref{Eq:required_ent_definition} under the correspondence $T_{max} \leftrightarrow p_{max}$, and $T \leftrightarrow p$. Here, we use $T_{max}=6$. As in Sec.~\ref{Section:balanced_bipartition_dataset1_QAOA}, we consider a volume-law scaling of the form ~\eqref{eq:fit}. 
This model has previously been used to describe entanglement scaling in 3-regular and weighted complete graphs under a linear annealing path \cite{DupontEntanglementPerspective2022}. Thus, it provides a reasonable starting point for characterizing entanglement in other MaxCut instances.

Here, we examine the robustness of the volume law under the different annealing schedules considered in this work, and compare the results with QAOA. Numerical simulations of entanglement scaling for $k$-regular graphs with $k=2,3,5$ and weighted complete graphs are shown in Fig.~\ref{Fig:volume_law_AQC}. First, we find that the volume law scaling persists across all schedules. Similarly, in QAOA, volume-law scaling is independent of the training strategy (see Sec~\ref{Section:balanced_bipartition_dataset1_QAOA}), suggesting a commonality between the underlying entanglement mechanisms of the two algorithms.  

Although the volume-law is maintained, the slope depends on graph properties and the annealing protocol. The optimized path generates less entanglement than the linear and QAOA-inspired paths, which is consistent with the behavior we observed in Fig.~\ref{Fig:Impact_Annealing_Path_AQC} for dense graphs. This is reasonable, as the optimized path is expected to navigate the solution space more effectively, avoiding highly entangled states when possible. 

Finally, we compare the entanglement requirements in AQC with the analytical benchmarks given by Eqs.~\eqref{Eq:PageRandom}  and \eqref{Eq:PageFermionicGaussianStates}. While numerical simulations for QAOA indicate that entanglement is bounded by that of FGSs (see Sec.~\ref{subsec:balanced-fgs}), this is not the case for AQC. As shown in Fig.~\ref{Fig:volume_law_AQC}, different graphs and annealing protocols can generate higher amounts of entanglement. These results demonstrate that, although QAOA and AQC are closely connected, entanglement scaling is sensitive to the specific components of the algorithms, even for instances within the same graph family.

\subsection{Entanglement Scaling with Graph Edge Density}

\begin{figure*}[t!]
    \centering
    \includegraphics[width=\textwidth]{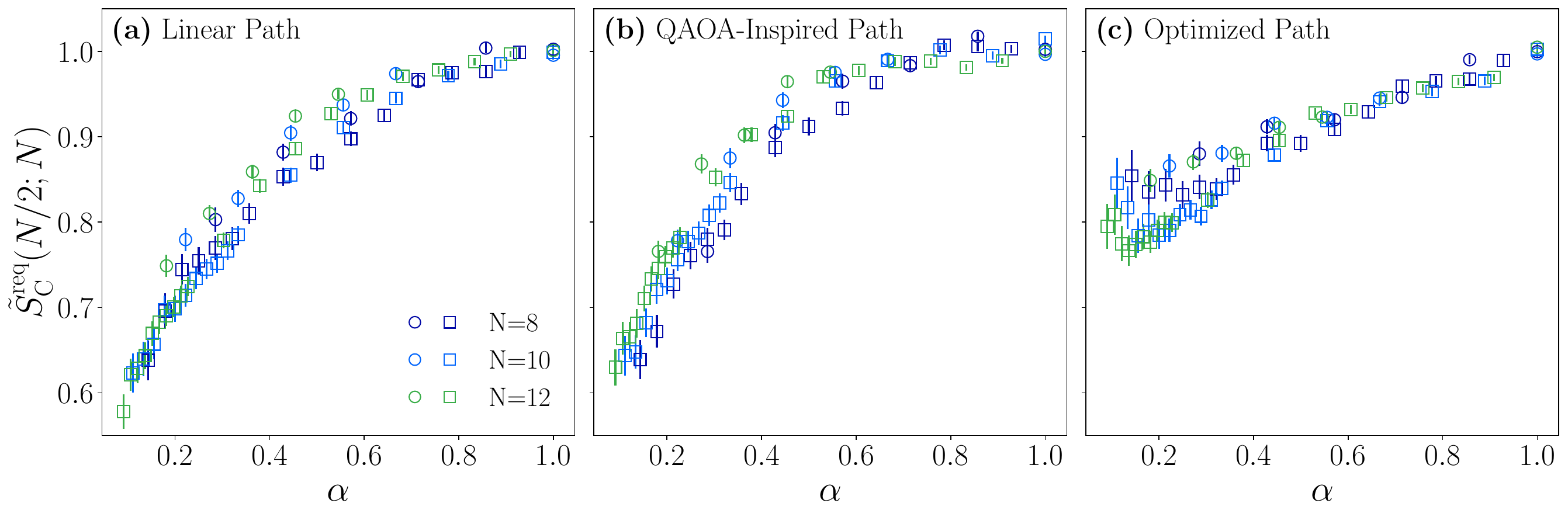}
    \caption{Scaling of $S^{\mathrm{req}}(N/2;N)$ in AQC for Dataset KR (circles) and Datasets CW, ED (squares) across multiple annealing schedules. The results are shown, normalized by complete graphs, as a function of edge density. Error bars denote the standard error of the mean.}  
    \label{Fig:volume_law_AQC_dataset2}
\end{figure*}

Thus far, our analysis of the entanglement scaling has focused on Datasets KR and CW. As discussed in Sec.~\ref{Section:balanced_bipartition_dataset1_dataset2_QAOA}, these datasets provide a controlled analysis based on the connectivity of each qubit, but at the same time they potentially limit the generality of the conclusions regarding AQC entanglement scaling. Here, we extend this study to Dataset ED, which contains graphs with varying number of edges. We incorporate results from this diverse set of graph instances to investigate the connection between AQC entanglement scaling and the graph's features.

We first examine entanglement scaling as a function of edge density for different annealing paths. The results for all three datasets are shown in  Fig.~\ref{Fig:volume_law_AQC_dataset2}(a)-(c), where values are normalized by the entanglement for complete graphs, similarly to Eq.~\eqref{Eq:normalized_Sreq_complete_fit}. The slope and intercept used for normalization differ across annealing paths, reflecting the linear fits shown in  Fig.~\ref{Fig:volume_law_AQC}. For the linear and QAOA-inspired paths, entanglement increases monotonically with edge density. However, the optimized path exhibits a non-monotonic behavior: entanglement initially decreases as a function of edge density up to a critical value  $\alpha = \alpha_c$, and then increases beyond this point. We attribute this distinct behavior in the low-edge density regime to the high degeneracy of the solution space. Because the optimized path is constructed to achieve a high approximation ratio, it tends to reach the solution space more efficiently, generating states whose entanglement is close to that of the true ground state. In contrast, the linear and QAOA-inspired paths do not reach the true ground state within the available evolution time $T=T_{max}$.
 
There is also a different scaling behavior in the large edge-density regime. For the optimized path, the scaling as a function of the edge density remains approximately linear,  consistent with the behavior observed in QAOA in Fig.~\ref{Fig:MaxEntGeneralGraphQAOA}. However, for linear and QAOA-inspired paths, the required entanglement gradually approaches that of complete graphs with a non-linear scaling. Despite these differences, data from different system sizes collapse onto a single curve for given choice of annealing path, suggesting that the entanglement requirement for arbitrary graphs can be predicted in all cases.

\section{Conclusions}
\label{Section:Summary}
We investigate the interplay between entanglement dynamics, algorithmic performance, and the features of the computational problem in QAOA. That is accomplished with noiseless simulations of the MaxCut problem across graphs with fixed and varying per-qubit connectivity. There are four important outcomes from this work. First, we show that the performance of the classical optimizer affects entanglement dynamics. Second, we investigate entanglement scaling with system size and edge density. In the former case, volume-law scaling is observed for both training strategies tested in this work. In the latter case, we present numerical evidence that entanglement scaling for an arbitrary graph can be predicted from its edge density (and system size). Third, we present numerical evidence connecting QAOA entanglement with FGSs. Finally, we consider entanglement dynamics and scaling in AQC with different annealing schedules. We show that, despite the close connection between QAOA and AQC, the two algorithms afford distinct entanglement scalings.

Our work paves the way for several directions for future research. First, the connection between QAOA states and FGSs identified in this work may motivate the design of novel QAOA ans\"atze that leverage qubit correlations to accelerate algorithmic training. Second, our conclusions regarding the impact of the classical optimizer on the entanglement profile should be extended to realistic noisy quantum devices. While noise typically degrades the performance of the classical optimizer, it also suppresses entanglement, reducing the required resources to solve the computational problem. Understanding the interplay between these competing effects can lead to nontrivial tradeoffs that merit further study. Finally, while our study focused on MaxCut, the relationship between entanglement scaling and problem structure should be examined for a broader class of problem Hamiltonians. For example, extending this study to more general Ising Hamiltonians is a possible next step, as many combinatorial optimization problems can be mapped onto this framework. 

\section{Acknowledgments}
This work was supported in part by the U.S. Department of Energy, Office of Science, Office of Advanced Scientific Computing Research under Award Number DE-SC0024163.

\appendix

\section{\texttt{INTERP} and Bilinear initialization strategy}
In Sec.~\ref{Section:OptimizationProcedure}, we detail the classical training strategy employed to update the variational parameters of QAOA. An important component of this procedure is the selection of an initial guess for the parameters, which can significantly influence the optimizer’s convergence and overall performance. In our approach, the initial guess is obtained using either the \texttt{INTERP} or the bilinear initialization protocol, with the method yielding the smaller cost function being selected.

\begin{figure}[t]
    \centering
    \includegraphics[width=\columnwidth,]{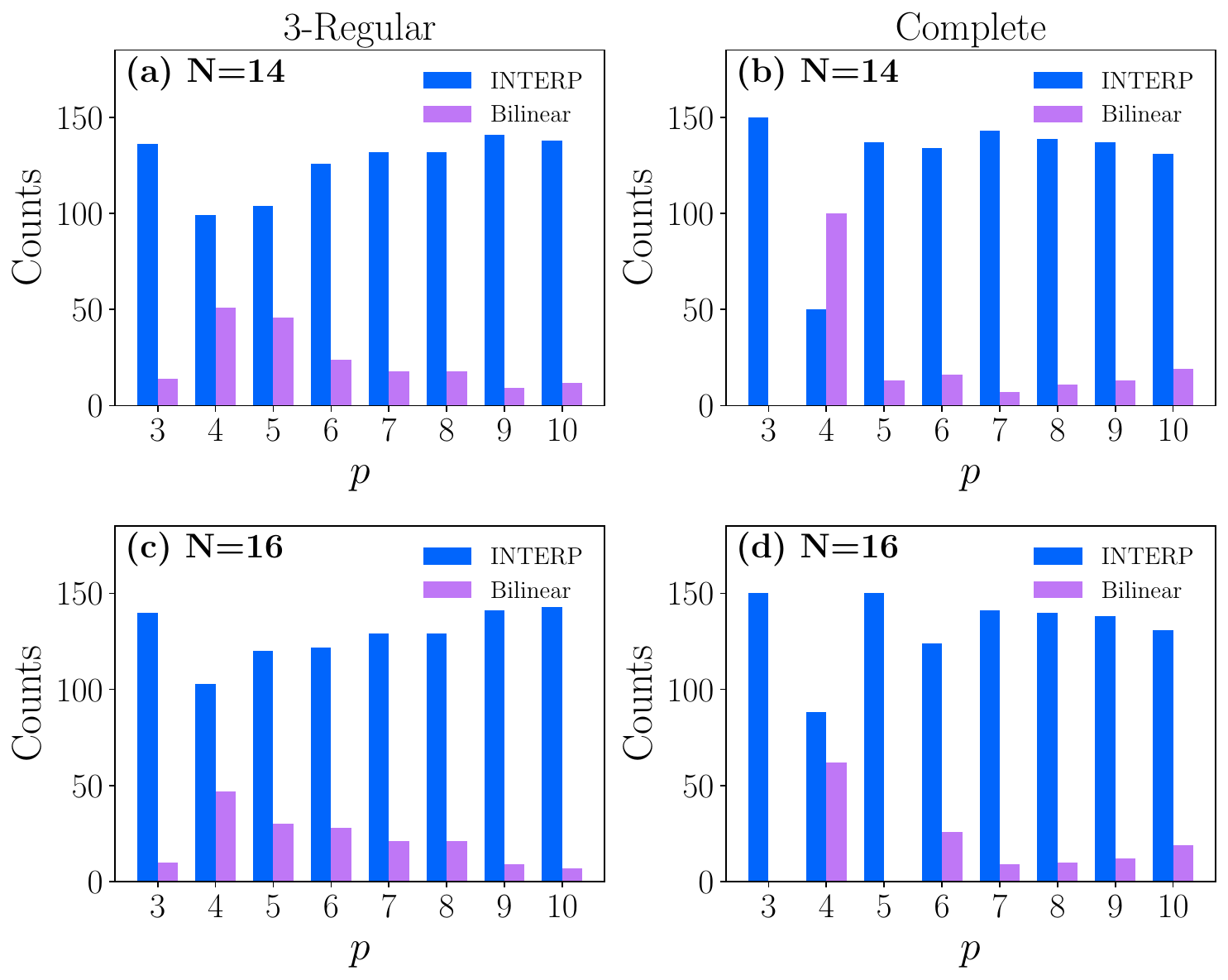}
    \caption{Number of instances in which the \texttt{INTERP} and bilinear initialization strategies are selected to initialize the optimization at different $p$-QAOA depths. Panels (a) and (c) show results for weighted 3-regular graphs with $N = 14$ and $N = 16$ qubits, respectively, while panels (b) and (d) present results for weighted complete graphs with the same system sizes. 150 graphs are considered for every case. The \texttt{INTERP} method typically generates initial guesses with lower cost values. The bilinear strategy also contributes significantly, particularly at $p = 4$ for complete graphs.}
    \label{Fig:INTERP_Bilinear}
\end{figure}

Here, we examine the frequency with which each initialization method is selected across different problem instances. The results are presented in Fig.~\ref{Fig:INTERP_Bilinear}, where we consider 3-regular and complete graphs with $N = 14, 16$ qubits.  For each $p$-QAOA instance, initial guesses are generated using both protocols, based on the optimized variational parameters from $(p{-}1)$- and $(p{-}2)$-QAOA. The histograms in Fig.~\ref{Fig:INTERP_Bilinear} display the selection frequency of each method, obtained by analyzing 150 graph instances from each graph family. 

\label{Section:INTERP_Bilinear_appendix}

\begin{figure}[t]
    \centering
    \includegraphics[width=\columnwidth,]{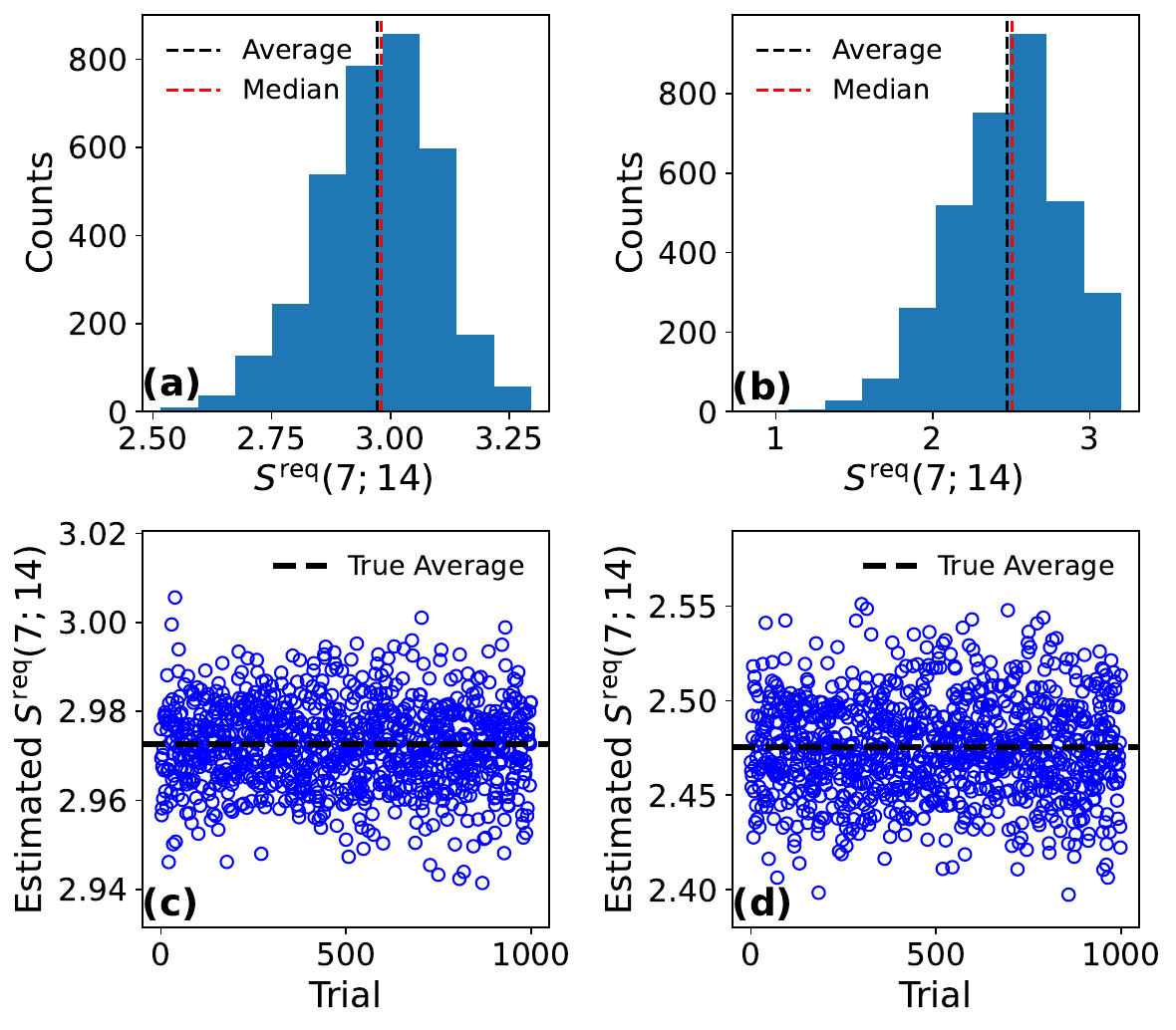}
    \caption{Evaluation of the sampling strategy for estimating $S^{\mathrm{req}}(N/2; N)$. Panels (a) and (b) show the distribution of $ S^{\mathrm{req}}(N/2; N)$ over all $\binom{N}{N/2}$ balanced bipartitions for a complete graph and a 3-regular graph, respectively. Panels (c) and (d) show the results from the sampling strategy over $10^3$ independent trials for the same graphs. In each trial, 5\% of all balanced bipartitions are randomly sampled, and the mean value is computed. Error bars (standard deviation $\approx 0.2$) are omitted for visual clarity. The sampling strategy closely matches the average over all $ \binom{N}{N/2}$ balanced bipartitions.}
    \label{Fig:Sampling_Strategy_Appendix}
\end{figure}

For both graph families, the \texttt{INTERP} method is more frequently selected, suggesting it often provides a more favorable initial cost. However, despite being chosen less frequently, the bilinear method yields lower cost values in a non-negligible number of instances. For complete graphs at depth $p = 4$, we observe a noticeable increase in the use of the bilinear strategy. Interestingly, this coincides with the QAOA depth where $S^{\mathrm{req}}(N/2;N)$ is observed in Fig.~\ref{Fig:entanglement_optimizer_AppendixB}(a), corresponding to the $p$-QAOA that generates the maximum entanglement. As discussed in Sec.~\ref{Section:optimization_strategy_effect_qaoa}, this entanglement peak for complete graphs has not been reported in previous works~\cite{DupontEntanglementPerspective2022,Usman2024CalibratingRoleOfEntanglement}, and we attribute this difference to the performance of the classical training protocol at this point. This analysis highlights the role of the initialization strategy in guiding the optimization through highly entangled and potentially rugged regions of the parameter space.

\section{Sampling strategy for entanglement calculation}
\label{Section:SamplingStrategyAppendix}

In this appendix, we describe and validate the sampling strategy used to estimate entanglement efficiently. In the main text, we analyze the behavior of entanglement dynamics across different settings in QAOA (and AQC). As the system size $N$ increases, the exact evaluation of $S^p_{\ell}(N_1; N)$, and subsequently of $S^p(N_1; N)$, and $S^{\text{req}}(N_1; N)$, becomes computationally demanding because these metrics must be averaged over multiple biparitions. Specifically, for a bipartition of an $N$-qubit system into subsystems containing $N_1$ and $N-N_1$ qubits, exact evaluation requires averaging over all $\binom{N}{N_1}$ possible configurations. In this work, we employ a sample-based strategy to estimate entanglement entropy efficiently.

We show that the sampling strategy produces accurate estimates of the entanglement, focusing here on $S^{\text{req}}(N/2; N)$. To this end, we consider two representative graph instances: a complete graph and a 3-regular graph, each with $N=14$ qubits. Other graph families and system sizes lead to similar conclusions. Panels (a) and (b) of Fig.~\ref{Fig:Sampling_Strategy_Appendix} show the distribution of $S^{\mathrm{req}}(N/2; N)$ across all bipartitions for these graph instances. They consist of the entanglement entropy values obtained from the full set of $\binom{N}{N/2}$ bipartitions. In both cases, the distributions exhibit concentrated peaks and low variance, as supported by the close agreement between the mean and median values. We leverage this result and, in this work,  $S^{\mathrm{req}}(N/2; N)$ is estimated using only 5\% of all possible bipartitions, rather than the full set required for the exact calculation. 

We assess the accuracy of this sampling-based approximation by benchmarking it against the exact average computed over all bipartitions. The corresponding analysis is presented in panels (c) and (d) of Fig.~\ref{Fig:Sampling_Strategy_Appendix}. Each point in these figures represents $S^{\mathrm{req}}(N/2; N)$ obtained by averaging over a random selection of 5\% of the bipartitions (approximately 172 samples for $N=14$). For each graph instance, this procedure is repeated over $10^3$ independent samplings. Across all trials and both graph classes, the sampled averages demonstrate strong agreement with the exact mean values. This result confirms that the reduced sampling scheme provides an accurate and computationally efficient estimator of the full average $S^{\mathrm{req}}(N/2; N)$, thereby enabling scalable analyses of larger quantum systems.

\end{document}